\documentclass[draft,tightenlines,nofootinbib,preprint,aps,eqsecnum,amsmath,amssymb]{revtex4}

\newcommand{\beq}{\begin{equation}}
\newcommand{\eeq}{\end{equation}}
\newcommand{\bea}{\begin{eqnarray}}
\newcommand{\eea}{\end{eqnarray}}

\newcommand{\sgn}{\mbox{\boldmath $\epsilon$}}

\newcommand{\byd}{\stackrel{def}{=}}

\begin{document}

\baselineskip 18pt

\today

\title{Charged Particles and the Electro-Magnetic Field in Non-Inertial Frames
of Minkowski Spacetime: II. Applications: Rotating Frames, Sagnac
Effect, Faraday Rotation, Wrap-up Effect}

\medskip

\author{David Alba}

\affiliation{
Sezione INFN di Firenze\\Polo Scientifico, via Sansone 1\\
 50019 Sesto Fiorentino, Italy\\
 E-mail ALBA@FI.INFN.IT}

\author{Luca Lusanna}

\affiliation{ Sezione INFN di Firenze\\ Polo Scientifico\\ Via Sansone 1\\
50019 Sesto Fiorentino (FI), Italy\\ Phone: 0039-055-4572334\\
FAX: 0039-055-4572364\\ E-mail: lusanna@fi.infn.it}

\begin{abstract}

We apply the theory of non-inertial frames in Minkowski space-time,
developed in the previous paper, to various relevant physical
systems. We give the 3+1 description without
coordinate-singularities of the rotating disk and the Sagnac effect,
with added  comments on pulsar magnetosphere and on a relativistic
extension of the Earth-fixed coordinate system. Then we study
properties of Maxwell equations in non-inertial frames like the
wrap-up effect and the Faraday rotation in astrophysics.

\end{abstract}

\maketitle

\vfill\eject

\section{Introduction}

In the first paper \cite{0} (quoted as paper I) we developed the
general theory of non-inertial frames in Minkowski space-time, whose
starting point are its admissible 3+1 splittings defining the
allowed conventions for clock synchronization, namely the allowed
notions of instantaneous 3-spaces needed, for instance, for setting
a well-posed Cauchy problem for Maxwell equations. In this way the
coordinate singularities of the traditional 1+3 approach are avoided
by construction. In particular it is shown that rigidly rotating
frames are not admissible in special relativity.

Also the formulation of charged particles and of the
electro-magnetic in non-inertial frames was given.

\bigskip

In this second paper we reformulate relevant physical system,
usually described in the 1+3 framework, in the non-inertial frames
based on the admissible 3+1 splittings.

\bigskip

In Section II there is a review of the rotating disk and of the
Sagnac effect in the 1+3 point of view followed by their description
in the framework of the 3+1 point of view (Subsection A) and by a
discussion on the ITRS rotating 3-coordinates fixed on the Earth
surface (Subsection B).\medskip

In Section III we give the 3+1 point of view in admissible nearly
rigidly rotating frames of the Wrap Up effect, of the Sagnac effect
and of the inertial Faraday rotation by studying electro-magnetic
wave solutions of the non-inertial Maxwell equations.
\medskip

In the Conclusions we give an overview of the results obtained in
these two papers and we identify the still open problems about
electro-magnetism in non-inertial frames.
\medskip

\vfill\eject

\section{The Rotating Disk and the Sagnac Effect}

In this Section we give the description of a rotating disk and of
the Sagnac effect starting from an admissible 3+1 splitting of
Minkowski space-time of the type of Eqs.(2.14) of I, i.e. whose
embedding has the form $z^{\mu}(\tau ,\sigma^u ) = x^{\mu}(\tau ) +
\epsilon^{\mu}_r\, R^r{}_s(\tau , \sigma )\, \sigma^s$ with
$x^{\mu}(\tau) = x^{\mu}_o + f^A(\tau)\, \epsilon^{\mu}_A$
describing the world-line of the observer origin of the
3-coordinates on the instantaneous 3-spaces $\Sigma_{\tau}$. This is
the simplest non-inertial frame whose 3-spaces are space-like
hyper-planes with admissible differentially rotating 3-coordinates.
The rotation matrix $R^r{}_s(\tau ,\sigma ) = R^r{}_s(\alpha_i(\tau,
\sigma )) = R^r{}_s(F(\sigma )\, {\tilde \alpha}_i(\tau))$ ($\sigma
= |\vec \sigma|$) is admissible if the function $F(\sigma)$
satisfies the M$\o$ller conditions $0 < F(\sigma ) < {1\over {A\,
\sigma}}$ and ${{d\, F(\sigma )}\over {d\sigma}} \not= 0$.

\medskip

An enlarged exposition of the material of this Section with a rich
bibliography is given in Section I Subsection D and E and in Section
VI Subsections B and C of the first paper in Ref.\cite{3}.

\bigskip

While at the non-relativistic level one can speak of a rigid (either
geometrical or material) disk put in global rigid rotatory motion,
the problem of the relativistic rotating disk is still under debate
(see Refs.\cite{16,43}) after one century from the enunciation of
the Ehrenfest paradox about the 3-geometry of the rotating disk. The
problems arise when one tries to define measurements of length, in
particular that of the circumference of the disk. Einstein \cite{44}
claims that while the rods along the radius $R_o$ are unchanged
those along the rim of the disk are Lorentz contracted: as a
consequence more of them are needed to measure the circumference,
which turns out to be greater than $2\pi\, R_o$ (non-Euclidean
3-geometry even if Minkowski space-time is 4-flat) and not smaller.
This was his reply to Ehrenfest \cite{45}, who had pointed an
inconsistency in the accepted special relativistic description of
the disk \footnote{If $R$ and $R_o$ denote the radius of the disk in
the rotating and inertial frame respectively, then we have $R = R_o$
because the velocity is orthogonal to the radius. But the
circumference of the rim of the disk is Lorentz contracted so that
$2\pi\, R < 2\pi\, R_o$ inconsistently with Euclidean geometry. } in
which it is the circumference to be Lorentz contracted: as a
consequence this fact was named the {\it Eherenfest paradox} (see
the historical paper of Gr$\o$n in Ref.\cite{46}).

\medskip

Since relativistic rigid bodies do not exist, at best we can speak
of {\it Born rigid motions} \cite{47} and {\it Born reference
frames} \footnote{A reference frame or platform is {\it Born-rigid}
\cite{48} if the expansion $\Theta$ and the shear $\sigma_{\mu\nu}$
of the associated congruence of time-like observers vanish, i.e. if
the spatial distance between neighboring world-lines remains
constant. }. However Gr$\o$n \cite{46} has shown that the
acceleration phase of a material disk is not compatible with Born
rigid motions and, moreover, we do not have a well formulated and
accepted relativistic framework to discuss a relativistic elastic
material disk.
\medskip

 As a consequence most of the authors treating the rotating disk
(either explicitly or implicitly) consider it as a {\it geometrical
entity} described by a congruence of time-like world-lines (helices
in Ref.\cite{49}) with non-zero vorticity, i.e. non-surface forming
and therefore non-synchronizable (see for instance Ref.\cite{50}).
This means that there is no notion of instantaneous 3-space where to
visualize the disk (see Ref.\cite{43} for the attempts to define
rods and clocks associated to this type of congruences): every
observer on one of these time-like world-lines can only define the
local rest frame and try to define a local accelerated reference
frame as said in Section IIB of paper I.\medskip

In the 3+1 point of view the disk is considered to be a relativistic
isolated system (either a relativistic material body or a
relativistic fluid or a relativistic dust as a limit case
\footnote{As an example of a congruence simulating a geometrical
rotating disk we can consider the relativistic dust described by
generalized Eulerian coordinates of Ref.\cite{51} after the gauge
fixing to a family of differentially rotating parallel
hyper-planes.}) with compact support always contained in a finite
time-like world-tube $W$, which in the Cartesian 4-coordinates of an
inertial system is a time-like cylinder of radius R. Each admissible
3+1 splitting of Minkowski space-time, centered on an arbitrary
time-like observer and with its two associated congruences of
time-like observers (see Section IIB of paper I), gives a
visualization of the disk in its instantaneous 3-spaces
$\Sigma_{\tau}$: at each instant $\tau$ the points of the disk in $W
\cap \Sigma_{\tau}$ are synchronized and through each one of them
pass an Eulerian observer belonging to the surface forming
congruence having as 4-velocity the unit normal to the instantaneous
3-spaces $\Sigma_{\tau}$. Instead the irrotational congruence of the
disk is described by the second congruence (whose unit 4-velocity is
$z^{\mu}_{\tau}(\tau ,\sigma^u )/ \sqrt{\sgn\, g_{\tau\tau}(\tau
,\sigma^u )}$  and whose observers follow generalized helices
$\sigma^u = \sigma^u_o$) associated to the admissible 3+1 splitting:
each of the observers of this congruence, whose world-lines are
inside $W$, has no intrinsic notion of synchronization.\medskip

As a consequence, each instantaneous 3-space $\Sigma_{\tau}$ of an
admissible 3+1 splitting has a well defined (in general Riemannian)
notion of 3-geometry and of spatial length: the radius and the
circumference of the disk are defined in $W \cap \Sigma_{\tau}$, so
that the disk 3-geometry is 3+1 splitting dependent. When the
material disk can be described by means of a parametrized Minkowski
theory, all these 3-geometry are gauge equivalent like the notions
of clock synchronization.\bigskip

The other important phenomenon connected with the rotating disk is
the {\it Sagnac effect} (see the recent review in Ref.\cite{52} for
how many interpretations of it exist), namely the phase difference
generated by the difference in the time needed for a round-trip by
two light rays, emitted in the same point, one co-rotating and the
other counter-rotating with the disk \footnote{For monochromatic
light in vacuum with wavelength $\lambda$ the fringe shift is
$\delta z = 4\, \vec \Omega \cdot \vec A / \lambda\, c$, where $\vec
\Omega $ is the Galilean velocity of the rotating disk supporting
the interferometer and $\vec A$ is the vector associated to the area
$|\vec A|$ enclosed by the light path. The time difference is
$\delta t = \lambda\, \delta z /c = 4\, \vec \Omega \cdot \vec
A/c^2$, which agrees, at the lowest order, with the proper time
difference $\delta \tau = (4\, A\, \Omega /c^2)\, (1 - \Omega^2\,
R^2/c^2)^{-1/2}$, $A = \pi\, R^2$, evaluated in an inertial system
with the standard rotating disk coordinates. This proper time
difference is twice the time lag due to the {\it synchronization
gap} predicted for a clock on the rim of the rotating disk with a
non-time orthogonal metric. See Refs.\cite{37,52,53} for more
details. See also Ref.\cite{36} for the corrections included in the
GPS protocol to allow the possibility of making the synchronization
of the entire system of ground-based and orbiting atomic clocks in a
reference local inertial system. Since  usually, also in GPS, the
rotating coordinate system has $t^{'} = t$ ($t$ is the time of an
inertial observer on the axis of the disk) the gap is a consequence
of the impossibility to extend Einstein's convention of the inertial
system also to the non-inertial one rotating with the disk: after
one period two nearby synchronized clocks on the rim are out of
synchrony.}. This effect, which has been tested (see the
bibliography of Refs.\cite{52,54}) for light, X rays and matter
waves (Cooper pairs, neutrons, electrons and atoms), has important
technological applications and must be taken into account for the
relativistic corrections to space navigation, has again an enormous
number of theoretical interpretations (both in special and general
relativity) like for the solutions of the Ehrenfest paradox. Here
the lack of a good notion of simultaneity leads to problems of {\it
time discontinuities or desynchronization effects} when comparing
clocks on the rim of the rotating disk.
\bigskip

Another area which is in a not well established form is
electrodynamics in non-inertial systems either in vacuum or in
material media ({\it problem of the non-inertial constitutive
equations}). Its clarification is needed both to derive the Sagnac
effect from Maxwell equations without gauge ambiguities \cite{37}
and to determine which types of experiments can be explained by
using the locality hypothesis (see Section IIB of paper I) to
evaluate the electro-magnetic fields in the comoving system (see the
Wilson experiment and the associated controversy \cite{40} on the
validity of the locality principle) without the need of a more
elaborate treatment like for the radiation of accelerated charges.
It would also help in the tests of the validity of special
relativity (for instance on the possible existence of a preferred
frame) based on Michelson-Morley - type experiments \cite{34,55}.

\bigskip

Instead (see also Ref.\cite{37}) we remark that the Sagnac effect
and the Foucault pendulum are {\it experiments which signal the
rotational non-inertiality of the frame}. The same is true for
neutron interferometry \cite{56}, where different settings of the
apparatus are used to {\it detect either rotational or translational
non-inertiality of the laboratory}. As a consequence a null result
of these experiments can be used to give a definition of {\it
relativistic quasi-inertial system}.

\bigskip

Let us remark that the disturbing aspects of rotations are rooted in
the fact that there is a deep difference between translations and
rotations at every level both in Newtonian mechanics and special
relativity:  the generators of translations satisfy an Abelian
algebra, while the rotational ones a non-Abelian algebra. As shown
in Refs.\cite{57}, at the Hamiltonian level we have that the
translation generators are the three components of the momentum,
while the generators of rotations are a pair of canonical variables
($L^3$ and $arctg\, {{L^2}\over {L^1}}$) and an unpaired variable
($|\vec L|$). As a consequence we can separate globally the motion
of the 3-center of mass of an isolated system from the relative
variables, but we cannot separate in a global and unique way three
Euler angles describing an overall rotation, because the residual
vibrational degrees of freedom are not uniquely defined.
\bigskip

We will now give the 3+1 point of view on these topics (Subsection
A), followed by a discussion on the rotating 3-coordinates fixed to
the Earth surface (Subsection B).

\subsection{The 3+1 Point of View on the Rotating Disk and
the Sagnac Effect.}

Let us describe an abstract geometrical disk with an admissible 3+1
splitting of the type (2.14) of I, in which the instantaneous
3-spaces are parallel space-like hyper-planes with normal $l^{\mu}$
centered on an inertial observer $x^{\mu}(\tau ) = l^{\mu}\, \tau$

\beq
 z^{\mu}(\tau ,\vec \sigma ) = l^{\mu}\, \tau +
 \epsilon^{\mu}_r\, R^r_{(3)\, s}(\tau ,\sigma )\,
 \sigma^s.
 \label{a1}
 \eeq

\medskip

The rotation matrix $R_{(3)}$ describes a differential rotation
around the fixed axis "3" (we take a constant $\omega$, but nothing
changes with $\omega(\tau )$)

 \bea
 &&R^r_{(3)\, s}(\tau ,\sigma ) = \left( \begin{array}{ccc} \cos\,
 \theta (\tau ,\sigma ) & - \sin\,  \theta (\tau ,\sigma )& 0\\
 \sin\,  \theta (\tau ,\sigma )& \cos\,  \theta (\tau ,\sigma )& 0\\
 0& 0& 1 \end{array} \right),\nonumber \\
 &&{}\nonumber \\
 &&\theta (\tau ,\sigma ) = F(\sigma )\, \omega\, \tau,\quad
 F(\sigma ) < {c\over {\omega\, \sigma}},\nonumber \\
 &&{}\nonumber \\
 &&\Omega^r{}_s(\tau ,\sigma ) = \left(R^{-1}_{(3)}\, {{d R_{(3)}}\over
 {d\tau}}\right){}^r{}_s(\tau ,\sigma ) = \omega\, F(\sigma )\,
 \left( \begin{array}{ccc} 0& -1& 0\\ 1& 0& 0\\ 0& 0& 0
 \end{array} \right),\nonumber\\
 &&\nonumber\\
 &&\Omega (\tau ,\sigma ) = \Omega (\sigma ) = \omega\,
 F(\sigma ).
 \label{a2}
 \eea

\bigskip

A simple choice for the gauge function $F(\sigma )$ is $F(\sigma ) =
{1\over {1 + {{\omega^2\, \sigma^2}\over {c^2}}}}$ (in the rest of
the Section we put $c = 1$), so that at spatial infinity we get
$\Omega (\tau ,\sigma ) = {{\omega}\over {1 + {{\omega^2\,
\sigma^2}\over {c^2}}}} \rightarrow_{\sigma \rightarrow \infty}\,
0$.

\bigskip

By introducing cylindrical 3-coordinates $r$, $\varphi$, $h$ by
means of the equations $\sigma^1 = r\, \cos\, \varphi$, $\sigma^2 =
r\, \sin\, \varphi$, $\sigma^3 = h$, $\sigma = \sqrt{r^2 + h^2}$, we
get the following form of the embedding and of its gradients

\begin{eqnarray*}
 z^{\mu}(\tau ,\vec \sigma ) &=& l^{\mu}\, \tau +
 \epsilon^{\mu}_1\, [\cos\, \theta (\tau ,\sigma )\, \sigma^1 -
 \sin\, \theta (\tau ,\sigma )\, \sigma^2] +\nonumber \\
 &&\nonumber\\
 &+& \epsilon^{\mu}_2\, [\sin\, \theta (\tau ,\sigma )\, \sigma^1 +
 \cos\, \theta (\tau ,\sigma )\, \sigma^2] + \epsilon^{\mu}_3\,
 \sigma^3 =\nonumber \\
&&\nonumber\\
 &=& l^{\mu}\, \tau + \epsilon^{\mu}_1\, r\, \cos\, [\theta (\tau
 ,\sigma ) + \varphi ] + \epsilon^{\mu}_2\, r\, \sin\, [\theta
 (\tau ,\sigma ) + \varphi ] + \epsilon^{\mu}_3\, h,
 \end{eqnarray*}

 \bea
  \frac{\partial
z^\mu(\tau,\vec{\sigma})}{\partial\tau}&=&
 z^{\mu}_{\tau}(\tau ,\vec \sigma ) = l^{\mu} - \omega\, r\,
 F(\sigma )\, \Big( \epsilon^{\mu}_1\, \sin\, [\theta
 (\tau ,\sigma ) + \varphi ] - \epsilon^{\mu}_2\, \cos\, [\theta
 (\tau ,\sigma ) + \varphi ]\Big),\nonumber \\
&&\nonumber\\
\frac{\partial z^\mu(\tau,\vec{\sigma})}{\partial\varphi}&=&
 z^\mu_\varphi(\tau,\vec{\sigma})=
-\epsilon^\mu_1\,r\,\sin\,[\theta(\tau,\sigma )+\varphi]
+\epsilon^\mu_2\,r\,\cos\,[\theta(\tau,\sigma )+\varphi]
\nonumber\\
&&\nonumber\\
\frac{\partial z^\mu(\tau,\vec{\sigma})}{\partial r}&=&
 z^\mu_{(r)}(\tau,\vec{\sigma})=
-\epsilon^\mu_1\,\left((\cos\,[\theta(\tau,\sigma )+\varphi]
-\frac{r^2\omega\tau}{\sqrt{r^2+h^2}}\,\frac{dF(\sigma)}{d\sigma}
\,\sin\,[\theta(\tau,\sigma )+\varphi]\right)+\nonumber\\
&&\nonumber\\
&+&\epsilon^\mu_2\,\left(\sin\,[\theta(\tau,\sigma )+\varphi]
+\frac{r^2\omega\tau}{\sqrt{r^2+h^2}}\,\cos\,[\theta(\tau,\sigma
)+\varphi]\right)
\nonumber\\
&&\nonumber\\
\frac{\partial z^\mu(\tau,\vec{\sigma})}{\partial h}&=&
 z^\mu_h(\tau,\vec{\sigma})=\epsilon^\mu_3
-\epsilon^\mu_1\,\left(\frac{rh\omega\tau}{\sqrt{r^2+h^2}}\,\frac{dF(\sigma)}{d\sigma}
\,\sin\,[\theta(\tau,\sigma )+\varphi]\right)+\nonumber\\
&&\nonumber\\
&+&\epsilon^\mu_2\,\left(\frac{rh\omega\tau}{\sqrt{r^2+h^2}}\,\frac{dF(\sigma)}{d\sigma}
\,\cos\,[\theta(\tau,\sigma )+\varphi]\right),
 \label{a3}
 \eea

\noindent where we have used the notation $(r)$ to avoid confusion
with the index $r$  used as 3-vector index (for example in
$\sigma^r$).
\bigskip

In the cylindrical 4-coordinates $\tau$, $r$, $\varphi$ and $h$ the
4-metric is

 \begin{eqnarray*}
 \sgn\,g_{\tau\tau}(\tau ,\vec \sigma ) &=& 1 - \omega^2\, r^2\,
 F^2(\sigma ),\qquad \sgn\,g_{\tau\varphi}(\tau ,\vec \sigma ) =
 -\omega\,r^2\,F(\sigma),\qquad \sgn\,g_{\varphi\varphi}(\tau ,\vec \sigma )
 = - r^2,\nonumber\\
 &&{}\nonumber\\
 \sgn\,g_{\tau (r)}(\tau ,\vec \sigma ) &=& -\frac{\omega^2\,r^3\,\tau}{\sqrt{r^2+h^2}}
\,F(\sigma)\,\frac{dF(\sigma)}{d\sigma},\qquad
 \sgn\,g_{\tau h}(\tau ,\vec \sigma ) = -\frac{\omega^2\,r^2\,h\,\tau}{\sqrt{r^2+h^2}}
\,F(\sigma)\,\frac{dF(\sigma)}{d\sigma},
 \end{eqnarray*}

\begin{eqnarray*}
 \sgn\,g_{(r)(r)}(\tau ,\vec \sigma ) &=&-1-\frac{r^4\,\omega^2\,\tau^2}{r^2+h^2}
\left(\frac{dF(\sigma)}{d\sigma}\right)^2,\nonumber\\
 &&\nonumber\\
 \sgn\,g_{hh}(\tau ,\vec \sigma ) &=&-1-\frac{r^2\,h^2\,\omega^2\,\tau^2}{r^2+h^2}
\left(\frac{dF(\sigma)}{d\sigma}\right)^2,\nonumber\\
 &&\nonumber\\
 \sgn\,g_{(r)\varphi}(\tau ,\vec \sigma ) &=&- \frac{\omega\,r^3\,\tau}{\sqrt{r^2+h^2}}
\,\frac{dF(\sigma)}{d\sigma},\qquad
 \sgn\,g_{h\varphi}(\tau ,\vec \sigma ) = - \frac{\omega^2\,r^2\,h\,\tau}{\sqrt{r^2+h^2}}
\,\frac{dF(\sigma)}{d\sigma},\nonumber\\
 &&\nonumber\\
 \sgn\,g_{h(r)}(\tau ,\vec \sigma ) &=& - \frac{r^3\,h\,\omega^2\,\tau^2}{r^2+h^2}
\left(\frac{dF(\sigma)}{d\sigma}\right)^2,
 \end{eqnarray*}

\bea
 &&with\, inverse\nonumber \\
 &&{}\nonumber \\
 \sgn\, g^{\tau\tau}(\tau ,\vec \sigma ) &=& 1,
 \qquad \sgn\, g^{\tau\varphi}(\tau ,\vec \sigma ) = - \omega\,
 F(\sigma ),\nonumber \\
 &&{}\nonumber \\
 \sgn\, g^{\tau (r)}(\tau ,\vec \sigma ) &=& \sgn\, g^{\tau h}(\tau ,\vec \sigma )
 = 0,\qquad \sgn\, g^{(r)(r)}(\tau ,\vec \sigma ) = \sgn\, g^{hh}(\tau ,\vec \sigma )
 = - 1,\nonumber \\
 &&{}\nonumber \\
 \sgn\, g^{\varphi\varphi}(\tau ,\vec \sigma ) &=& - {{1 + \omega^2\, r^2\, [\tau^2\,
 ({{dF(\sigma )}\over {d\sigma}})^2 - F^2(\sigma )}\over
 {r^2}},\nonumber \\
 &&{}\nonumber \\
 \sgn\, g^{\varphi (r)}(\tau ,\vec \sigma ) &=& {{\omega\, r\,
 \tau}\over {\sqrt{r^2 + h^2}}}\, {{dF(\sigma )}\over
 {d\sigma}},\qquad \sgn\, g^{\varphi h}(\tau ,\vec \sigma ) = {{\omega\, h\,
 \tau}\over {\sqrt{r^2 + h^2}}}\, {{dF(\sigma )}\over
 {d\sigma}}.
 \label{a4}
 \eea
\bigskip

It is easy to observe that the congruence of (non inertial)
observers defined by the 4-velocity field

 \beq
  {{z^{\mu}_\tau(\tau
,\vec \sigma )}\over {\sqrt{\sgn\, g_{\tau\tau}(\tau ,\vec \sigma
)}}} = {{l^{\mu} - \omega\, r\, F(\sigma )\, \Big(
\epsilon^{\mu}_1\, \sin\, [\theta
 (\tau ,\sigma ) + \varphi ] - \epsilon^{\mu}_2\, \cos\, [\theta
 (\tau ,\sigma ) + \varphi ]\Big)}\over
{1 - \omega^2\, r^2\, F^2(\sigma )}},
 \label{a5}
  \eeq

\noindent has the observers moving along the world-lines

 \bea
 &&x^{\mu}_{{\vec\sigma}_o}(\tau ) = z^{\mu}(\tau ,{\vec \sigma}_o)
= \nonumber\\ &&\nonumber\\ &=&l^{\mu}\, \tau + r_o\, \Big(
\epsilon^{\mu}_1\, \cos\, [\omega\, \tau\, F(\sigma_o) + \varphi_o]
+ \epsilon^{\mu}_2\, \sin\, [\omega\, \tau\, F(\sigma_o) +
\varphi_o]\Big) + \epsilon^{\mu}_3\, h_o.
 \label{a6}
  \eea

The world-lines (\ref{a6}) are labeled by their initial value
$\vec{\sigma} = {\vec \sigma}_o = (\varphi_o,r_o,h_o)$ at $\tau =
0$.
\bigskip

In particular for $h_o=0$ and $r_o=R$ these world-lines are {\it
helices} on the {\em  cylinder} in the Minkowski space

 \beq
 \epsilon_3^\mu\,z_\mu = 0,\qquad
\left(\epsilon_1^\mu\,z_\mu\right)^2 +
\left(\epsilon_2^\mu\,z_\mu\right)^2 = R^2,
 \qquad or \qquad r = R, \qquad h = 0.
   \label{a7}
    \eeq

These helices are defined the equations $\varphi = \varphi_o,\, r =
R,\,h = 0$ if expressed in the embedding adapted coordinates
$\varphi,r,h$. Then the congruence of observers (\ref{a5}), defined
by the foliation (\ref{a1}), defines on the cylinder (\ref{a7}) the
{\em rotating observers} usually assigned to the rim of a rotating
disk, namely  observes running along the helices $
x^{\mu}_{{\vec\sigma}_o}(\tau ) = l^{\mu}\, \tau + R\, \Big(
\epsilon^{\mu}_1\, \cos\, [\Omega(R)\, \tau\, + \varphi_o] +
\epsilon^{\mu}_2\, \sin\, [\Omega(R)\, \tau\,  + \varphi_o]\Big)$
after having put $\Omega(R)\equiv\omega\,F(R)$.
\bigskip

On the cylinder (\ref{a7}) the line element is obtained from the
line element $ds^2 = g_{AB}\, d\sigma^A\, d\sigma^B$ for the metric
(\ref{a4}) by putting $dh = dr = 0$ and $r = R$, $h = 0$. Therefore
the cylinder line element is

 \beq
  \sgn\,(ds_{cyl})^2 = \Big[1 - \omega^2\, R^2\,
F^2(R)\Big]\, (d\tau )^2 - 2\, \omega\, R^2\, F(R)\, d\tau d\varphi
- R^2\, (d\varphi )^2.
 \label{a8}
  \eeq
\medskip

We can define the {\it light rays on the cylinder}, i.e. the null
curves on it, by solving the equation

\begin{equation}
 \sgn\,(ds_{cyl})^2=(1-R^2\,\Omega^2(R))\,d\tau^2-2\,R^2\,\Omega(R)\,d\tau\,d\varphi-
R^2\,d\varphi^2=0,
 \label{a9}
 \end{equation}

\noindent which implies

\begin{equation}
 R^2\,\left(\,\frac{d\varphi(\tau)}{d\tau}\,\right)^2+
2\,R^2\,\Omega(R)\,\left(\,\frac{d\varphi(\tau)}{d\tau}\,\right)-
(1-R^2\,\Omega(R))=0.
 \label{a10}
 \end{equation}

\bigskip

The two solutions

\begin{equation}
\frac{d\varphi(\tau)}{d\tau}=\pm\,\frac{1}{R}-\Omega(R),
 \label{a11}
 \end{equation}

\noindent define the world-lines on the cylinder for {\em clockwise
or anti-clockwise} rays of light.

\begin{equation}
 \begin{array}{l}
\Gamma_1:\qquad \varphi (\tau ) - \varphi_o =
\left(+\frac{1}{R}-\Omega(R)\right)\,\tau ,\\ \\
 \Gamma_2:\qquad
\varphi (\tau ) - \varphi_o =
\left(-\frac{1}{R}-\Omega(R)\right)\,\tau
\end{array}.
 \label{a12}
 \end{equation}

\bigskip

This is the {\em geometric origin} of the {\em Sagnac effect}. Since
$\Gamma_1$ describes the world-line of the ray of light emitted at
$\tau=0$ by the rotating observer $\varphi=\varphi_o$ in the
increasing sense of $\varphi$ (anti-clockwise), while $\Gamma_2$
describes that of the ray of light emitted at $\tau=0$ by the same
observer in the decreasing sense of $\varphi$ (clockwise), then the
two rays of light will be re-absorbed by the same observer at {\it
different $\tau$-times} \footnote{Sometimes the {\em proper time of
the rotating observer} is used: $d{\cal
T}_o=d\tau\sqrt{1-\Omega^2(R)\,R^2}$.} $\tau_{(\pm\, 2\pi ) }$,
whose value, determined by the two conditions $\varphi (\tau_{(\pm\,
2\pi )}) - \varphi_o= \pm\, 2\pi$, is

\beq
 \begin{array}{l}
\Gamma_1:\qquad \tau_{(+2\pi )} =
\frac{2\pi\,R}{1-\Omega(R)\,R},\qquad \Gamma_2:\qquad \tau_{(-2\pi
)}=\frac{2\pi\,R}{1+\Omega(R)\,R}.
\end{array}
 \label{a13}
 \eeq

The {\it time difference} between the re-absorption of the two light
rays is\medskip

\begin{equation}
\Delta\tau = \tau_{(+2\pi )} - \tau_{(-2\pi)} =
\frac{4\pi\,R^2\,\Omega(R)}{1-\Omega^2(R)\,R^2} =
\frac{4\pi\,R^2\,\omega\, F(R)}{1-\omega^2\, F^2(R)\,R^2},
 \label{a14}
 \end{equation}

\noindent and it corresponds to the phase difference named the {\em
Sagnac effect} (see footnote 4)

\begin{equation}
 \Delta\Phi=\Omega\,\Delta\tau,\qquad \Omega = \Omega (R) = \omega\, F(R) .
 \label{a15}
 \end{equation}
\medskip

We see that we recover the standard result if we take a function
$F(\sigma )$ such that $F(R) = 1$. In the non-relativistic
applications, where $F(\sigma ) \rightarrow 1$, the correction
implied by admissible relativistic coordinates is totally
irrelevant.

\bigskip

With an admissible notion of simultaneity, all the clocks on the rim
of the rotating disk lying on a hyper-surface $\Sigma_{\tau}$ are
automatically synchronized. Instead for rotating observers of the
irrotational congruence there is a {\em desynchronization effect or
synchronization gap} because they cannot make a global
synchronization of their clocks: {\it usually a discontinuity in the
synchronization of clocks is accepted and taken into account} (see
Ref.\cite{36} for the GPS).
\medskip

To clarify this point and see the emergence of this gap, let us
consider a reference observer $(\varphi_o = const., \tau )$ and
another one $(\varphi = const. \not= \varphi_o, \tau )$. If $\varphi
> \varphi_o$ we use the notation $(\varphi_R, \tau )$, while for
$\varphi < \varphi_o$ the notation $(\varphi_L, \tau )$ with
$\varphi_R - \varphi_o = - (\varphi_L - \varphi_o)$.

Let us consider the two rays of light $\Gamma_{R\, -}$ and
$\Gamma_{L\, -}$, with world-lines given by Eqs.(\ref{a12}), emitted
in the right and left directions at the event $(\varphi_o,\tau_{-})$
on the rim of the disk and received at $\tau$ at the events
$(\varphi_R,\tau)$ and $(\varphi_L,\tau )$ respectively. Both of
them are  reflected towards the reference observer, so that we have
two rays of light $\Gamma_{R\, +}$ and $\Gamma_{L\, +}$ which will
be absorbed at the event $(\varphi_o,\tau_{+})$. By using
Eq.(\ref{a13}) for the light propagation, we get

\bea
 && \Gamma_{R\, -}:\,\,  (\varphi  -
\varphi_o) = \frac{1-R\Omega(R)}{R}\,(\tau - \tau_{-}),\qquad
\Gamma_{R\, +}:\,\, (\varphi  - \varphi_o) =
\frac{1+R\Omega(R)}{R}\,(\tau_{+} - \tau),\nonumber \\
 &&{}\nonumber \\
 &&\Gamma_{L\, -}:\,\, (\varphi  - \varphi_o) =
-\frac{1+R\Omega(R)}{R}\,(\tau - \tau_{-}),\qquad \Gamma_{L\,
+}:\,\, (\varphi  - \varphi_o) = -\frac{1-R\Omega(R)}{R}\,(\tau_{+}
- \tau).\nonumber \\
 &&{}
 \label{a16}
 \eea

\medskip

As shown in Section II, Eqs.(2.17) and (2.18), of the first paper in
Ref.\cite{3}, in a neighborhood of the observer $(\varphi_o ,\tau )$
[$(\varphi ,\tau )$ is an observer in the neighborhood] we can only
define the following local synchronization \footnote{See
Ref.\cite{53} for a derivation of the Sagnac effect in an inertial
system by using Einstein's synchronization in the locally comoving
inertial frames on the rim of the disk and by asking for the
equality of the one-way velocities in opposite directions.}

\begin{equation}
 c\, \Delta\,\widetilde{\cal T}= \sqrt{1-R^2\,\Omega^2(R)}\,\Delta
\tau_E= \sqrt{1-R^2\,\Omega^2(R)}\,\Delta \tau-
\frac{R^2\,\Omega^2(R)}{\sqrt{1-R^2\Omega^2(R)}} \,\Delta \varphi.
 \label{a17}
 \end{equation}
\medskip

If we try to extend this local synchronization  to a global one for
two distant observers $(\varphi_o, \tau )$ and $(\varphi ,\tau )$ in
the form of an Einstein convention  (the result is the same both for
$\varphi = \varphi_R$ and $\varphi = \varphi_L$)

\begin{equation}
\tau_E=\frac{1}{2}\,(\tau_{+} + \tau_{-}) = \tau-
\frac{R^2\Omega(R)}{1-R^2\,\Omega^2(R)}(\varphi  - \varphi_o),
 \label{a18}
 \end{equation}

\noindent we arrive at a contradiction, because the curves defined
by $\tau_E=constant$ are {\it not closed}, since they are helices
that assign the {\it same time} $\tau_E$ to different events on the
world-line of an observer $\varphi_o=constant$. For example
$(\varphi_o,\tau)$ and $\left(\varphi_o,\tau+2\pi
\,\frac{R^2\Omega(R)}{1-R^2\,\Omega^2(R)}\right)$ are on the same
helix $\tau_E=constant$. As a consequence we get the synchronization
gap.

\bigskip

As shown in both papers of Ref.\cite{3}, by using the global
synchronization on the instantaneous 3-spaces $\Sigma_{\tau}$ we can
define a generalization of Einstein's convention for clock
synchronization by using the radar time $\tau$. If an accelerated
observer A emits a light signal at $\tau_{-}$, which is reflected at
a point P of the world-line of a second observer B and then
reabsorbed at $\tau_+$, then the B clock at P has to be synchronized
with the following instant of the A clock [$n = +$ for $\varphi =
\varphi_R$, $n = -$ for $\varphi = \varphi_L$]

\bea
 \tau(\tau_{-}, n, \tau_{+}) &=&
\frac{1}{2}(\tau_{+} + \tau_{-}) - \frac{n\,R\,\Omega(R)}{2}\,
(\tau_{+} - \tau_{-})\, {\buildrel {def}\over =}\, \tau_{-} +
 {\cal E}(\tau_{-}, n, \tau_{+})\, (\tau_{+} - \tau_{-}),\nonumber \\
 &&{}\nonumber \\
 with &&\qquad {\cal E}(\tau_{-}, n, \tau_{+}) = {{1 - n\, R\,
 \Omega (R)}\over 2},\qquad \Omega (R) = \omega\, F(R).
 \label{a19}
 \eea
\bigskip

Finally in the first paper of Ref.\cite{3} [see Eqs.(6.37)-(6.47) of
Section VI] there is the evaluation of the radius and the
circumference of the rotating disk. If we choose the spatial length
of the instantaneous 3-space $\Sigma_{\tau}$ of the admissible
embedding (\ref{a1}), we get an Euclidean 3-geometry, i.e. a
circumference $2\pi\, R$ and a radius $R$ at each instant $\tau$
independently from the choice of the gauge function $F(\sigma )$.
With other admissible 3+1 splittings we would get non-Euclidean
results: as said they are gauge equivalent when the disk can be
described with a parametrized Minkowski theory. Instead the use of a
local notion of synchronization from the observers of the
irrotational congruence located on the rim of the rotating disk
implies a local definition of spatial distance based on the 3-metric
${}^3\gamma_{uv} = - \sgn\, \Big(g_{uv} - \frac{g_{\tau u}\,g_{\tau
v}}{g_{\tau\tau}}\Big)$, i.e. a non-Euclidean 3-geometry. In this
case the radius is $R$, but the circumference is $2\pi\, R/ \sqrt{1
- R^2\, \Omega^2(R)}$. However this result holds only in the local
rest frame of the observer with the tangent plane orthogonal to the
observer 4-velocity (also called the abstract relative space)
identified with a 3-space (see Section IIB of paper I).
\bigskip

See Subsection D of Section III for a derivation of the Sagnac
effect in nearly rigid rotating frames.

\subsection{The Rotating ITRS 3-Coordinates fixed on the Earth Surface.}

The embedding (2.14) of I, describing admissible differential
rotations in an Euclidean 3-space, could be used to improve the
conventions IERS2003 (International Earth Rotation and Reference
System Service) \cite{39} on the non-relativistic transformation
from the 4-coordinates of the Geocentric Celestial Reference System
(GCRS) to the International Terrestrial Reference System (ITRS), the
Earth-fixed geodetic system of the new theory of Earth rotation
replacing the old precession-nutation theory. It would be a special
relativistic improvement to be considered as an intermediate step
till to a future development leading to a post-Newtonian (PN)
general relativistic approach unifying the existing non-relativistic
theory of the geo-potential below the Earth surface with the GCRS PN
description of the geo-potential outside the Earth given by the
conventions IAU2000 (International Astronomical Union) \cite{39} for
Astrometry, Celestial Mechanics and Metrology in the relativistic
framework.
\bigskip

In the IAU 2000 Conventions the Solar System is described in the
Barycentric Celestial Reference System (BCRS) as a {\it
quasi-inertial} frame, centered on the barycenter of the Solar
System, with respect to the Galaxy. BCRS is parametrized with
harmonic PN 4-coordinates $x^{\mu}_{BCRS} = \Big(x^o_{BCRS} = c\,
t_{BCRS}; x^i_{BCRS}\Big)$, where $t_{BCRS}$ is the barycentric
coordinate time and the mutually orthogonal spatial axes are {\it
kinematically non-rotating} with respect to fixed radio sources.
This a nearly Cartesian 4-coordinate system in a PN Einstein
space-time and there is an assigned 4-metric, determined modulo
$O(c^{-4})$ terms and containing the gravitational potentials of the
Sun and of the planets, PN solution of Einstein equations in
harmonic gauges: in practice it is considered as a special
relativistic inertial frame with nearly Euclidean instantaneous
3-spaces $t_{BCRS} = const.$ (modulo $O(c^{-2})$ deviations) and
with Cartesian 3-coordinates $x^i_{BCRS}$. This frame is used for
space navigation in the Solar System. The geo-center (a fictitious
observer at the center of the earth geoid) has a world-line
$y^{\mu}_{BCRS}(x^o_{BCRS}) = \Big(x^o_{BCRS};
y^i_{BCRS}(x^o_{BCRS})\Big)$, which is approximately a straight
line.

\bigskip

For space navigation near the Earth (for the Space Station and near
Earth satellites using NASA coordinates) and for the studies from
spaces of the geo-potential one uses the GCRS, which is defined
outside the Earth surface as a local reference system centered on
the geo-center. Due to the earth rotation of the Earth around the
Sun, it deviates from a nearly inertial special relativistic frame
on time scales of the order of the revolution time. Its harmonic
4-coordinates $x^{\mu}_{GCRS} = \Big(x^o_{GCRS} = c\, t_{GCRS};
x^i_{GCRS}\Big)$, where $t_{GCRS}$ is the geocentric coordinate
time, are obtained from the BCRS ones by means of a PN coordinate
transformation which may be described as a special relativistic pure
Lorentz boost without rotations (the parameter is the 3-velocity of
the geo-center considered constant on small time scales) plus
$O(c^{-4})$ corrections taking into account the gravitational
acceleration of the geo-center induced by the Sun and the planets.
As a consequence the GCRS spatial axes are kinematically
non-rotating in BCRS and the relativistic inertial forces (for
instance the Coriolis ones) are hidden in the geodetic precession;
the same holds for the aberration effects and the dependence on
angular variables. A PN 4-metric, determined modulo $O(c^{-4})$
terms, is given in IAU2000: it also contains the GCRS form of the
geo-potential and the inertial and tidal effects of the Sun and of
the planets. Again the instantaneous 3-spaces are considered nearly
Euclidean (modulo $O(c^{-2})$ deviations) 3-spaces $t_{GCRS} =
const.$.\bigskip

In IAU200 the coordinate times $t_{BCRS}$ and $t_{GCRS}$ are then
connected with the time scales used on Earth: SI Atomic Second, TAI
(International Atomic Time), TT (Terrestrial Time), $T_{EPH}$
(Ephemerides Time), UT and UT1 and UTC (Universal Times for civil
use), GPS (Mastr Time), ST (Station Time).\bigskip

Finally we need a 4-coordinate system fixed on the Earth crust. It
is the ITRS with 4-coordinates $x^{\mu}_{ITRS} = \Big(x^o_{ItRS}
{\buildrel {def}\over =}\, c\, t_{GCRS}; x^i_{ITRS}\Big)$, which
uses the same coordinate time as GCRS. It is obtained from GCRS by
making a suitable set of non-relativistic time-dependent {\it rigid}
rotations on the nearly Euclidean 3-spaces $t_{GCRS} = const.$. The
geocentric rectangular 3-coordinates $x^i_{ITRS}$ match the
reference ellipsoid WGS-84 (basis of the terrestrial coordinates
(latitude, longitude, height) obtainable from GPS) used in geodesy.
As shown in IERS2003, we have $x^i_{ITRS} = \Big( W^T(t_{GCRS})\,
R^T_3(- \theta )\, C \Big)^i{}_j\, x^j_{GCRS}$, where $C =
R^T_3(s)\, R^T_3(E)\, R^T_2(- d)\, R^T_3(- E)$ and $W(t_{GCRS}) =
R_3(- s^{'})\, R_2(x_p)\, R_1(y_p)$ are rotation matrices. This
convention is based on the new definition of the Earth rotation axis
($\theta$ is the angle of rotation about this axis): it is the line
through the geo-center in direction of the Celestial Intermediate
Pole (CIP) at date $t_{GCRS}$, whose position in GCRS is $n^i_{GCRS}
= \Big(sin\, d\, cos\, E, sin\, d\, sin\, E, cos\, d\Big)$. The new
non-rotating origin (NLO) of the rotation angle $\theta$ on the
Earth equator (orthogonal to the rotation axis) is a point named the
Celestial Intermediate Origin (CIO), whose position in CGRS requires
the angle $s$, called the CIO locator. Finally in the rotation
matrix $W^T(t_{GCRS})$ (named the polar motion or wobble matrix) the
angles $x_p$ and $y_p$ are the angular coordinates of CIP in ITRS,
while the angle $s^{'}$ is connected with the re-orientation of the
pole from the ITRS z-axis to the CIP plus a motion of the origin of
longitude from the ITRS x-axis to the so-called Terrestrial
Intermediate Origin (TIO), used as origin of the azimuthal angle.

\bigskip

Let us now consider the embedding $z^{\mu}(\tau ,\sigma^u) =
x^{\mu}(\tau) + \epsilon^{\mu}_r\, R^r{}_s(\tau ,\sigma )\,
\sigma^s$ of Eq.(2.14) of I. Let us identify $x^{\mu} = z^{\mu}(\tau
,\sigma^u)$ with the GCRS 4-coordinates $x^{\mu}_{GCRS}$ centered on
the world-line of the geo-center assumed to move along a straight
line. Then, if we identify the space-like vectors $\epsilon^{\mu}_r$
with the GCRS non-rotating spatial axes, we have $x^{\mu}(\tau ) =
\epsilon^{\mu}_{\tau}\, \tau = l^{\mu}\, \tau$, where $l^{\mu}$ is
orthogonal to the nearly Euclidean 3-spaces $t_{GCRS} = const.$. The
proper time $\tau$ of the geo-center coincides with $c\, t_{GCRS}$
modulo $O(c^{-2})$ corrections from the GCRS PN 4-metric.

Then a special relativistic definition of ITRS can be done by
replacing the rigidly rotating 3-coordinates $x^i_{ITRS}$ with the
differentially rotating 3-coordinates $\sigma^r$. The rotation
matrix $R(\tau ,\sigma )$, with the choice $F(\sigma ) = {1\over {1
+ {{\omega^2\, \sigma^2}\over {c^2}}}}$ for the gauge function
($\omega$ can be taken equal to the mean angular velocity for the
Earth rotation), will contain three Euler angles determined by
putting $R(\tau ,\sigma ){|}_{F(\sigma) = 1} = C^T\, R_3(- \theta)\,
W(t_{GCRS})$.

In this way a special relativistic version of ITRS could be given as
a preliminary step towards a PN general relativistic formulation of
the geo-potential inside the Earth to be joined consistently with
GCRS outside the Earth. Even if this is irrelevant for geodesy
inside the geoid, it could lead to a refined treatment of effects
like geodesic precession taking into account a model of
geo-potential interpolating smoothly between inside and outside the
geoid and the future theory of heights over the reference ellipsoid
under development in a formulation of relativistic geodesy based on
the use of  the new generation of microwave and optical atomic
clocks both on the Earth surface and in space.

\vfill\eject

\section{Non-Inertial Maxwell Equations in Nearly Rigid Rotating Frames}

In the 3+1 point of view the Maxwell equations (4.17) of I in an
arbitrary inertial frame are identical to the Maxwell equations in
general relativity, but now the 4-metric is describing only the
inertial effects present in the given frame. Therefore we can adapt
the techniques used in general relativity to non-inertial frames,
for instance the definition of electric and magnetic fields done in
Ref.\cite{28} (see Appendix A of paper I) or the geometrical optic
approximation to light rays of Ref.\cite{29}.

For the 1+3 point of view on this topic see for instance
Ref.\cite{30} and its bibliography. In particular, for the treatment
of electromagnetic wave in rotating frame by means of Fermi
coordinates \cite{31} and for the determination of the
helicity-rotation coupling, as a special case of spin-rotation
coupling \cite{32,33}. In all these calculations the locality
hypothesis (see Section IIB of paper I) is used.

In the case of linear acceleration an analysis of the inertial
effects has been done in Ref.\cite{34}. The same non-inertial
4-metric has been used in Ref.\cite{35} to study the optical
position meters constituents of the laser interferometers on ground
used for the detection of gravitational waves. However the 4-metric
used has a bad behavior at spatial infinity, so that the conclusions
on the electro-magnetic waves in these frames (even if supposed to
hold at distances smaller than those where there are coordinate
singularities) are questionable because the Cauchy problem for
Maxwell equations is not well posed.

\medskip

In this Section we study some properties of electro-magnetic waves
and of geometrical optic approximation to light rays in the
radiation gauge in the admissible rotating non-inertial frame
defined by the embedding (2.14) of I, ensuring a well-posed Cauchy
problem, at small distances from the rotation axis where the
$O(c^{-1})$ deviations from rigid rotations is governed by
Eqs.(2.15) and (2.16) of I. Even if we will ignore these deviations,
doing the calculations in the radiation gauge in locally rigidly
rotating frames, they could be taken into account in a more refined
version of the subsequent calculations base on the 3+1 point of
view, which is free from coordinate singularities. This would also
allow to verify the validity of the locality hypothesis. In
particular we consider the Phase Wrap Up effect \cite{31,36}, the
Sagnac effect \cite{37,31a} and the Faraday Rotation \cite{38}.

\subsection{The 3+1 Point of View on Electro-Magnetic Waves and Light Rays  in
Nearly Rigidly Rotating Non-Inertial Frames.}

Let us consider a non-inertial frame of the type (2.14) of I with
vanishing linear acceleration and $\tau$-independent angular
velocity and centered on an inertial observer. In the notation of
Eqs.(2.15), (2.16) and (4.47) of I, we have $x^{\mu}(\tau) =
\epsilon^{\mu}_{\tau}\, \tau$, i.e. $\vec v(\tau ) = \vec w(\tau) =
0$, and $\vec \Omega (\tau) = \vec \Omega = const.$ (whose
components are ${\tilde \Omega}^r = const.$). We will ignore the
higher order terms, so that locally we have a rigidly rotating
frame, but with more effort small deviations from rigid rotation
could be taken into account.

\medskip

In this case the Hamiltonian (4.35), or (4.51), of I gives the
following Hamilton equations for the transverse electro-magnetic
field (${\vec A}_{\perp} = \{A_{\perp\, r} = {\tilde A}^r_{\perp}
\not= A^r_{\perp} \} + O(c^{-2})$)

 \bea
\frac{\partial {\tilde
A}^r_\perp(\tau,\vec{\sigma})}{\partial\tau}&=&
\pi^r_\perp(\tau,\vec{\sigma}) -
 {1\over c}\, \int d^3\sigma'\, \left[ - \vec{\Omega} \cdot
\vec{\sigma}' \times {\vec \partial}'\,
\vec{A}_\perp(\tau,\vec{\sigma}') + \vec{\Omega} \times
\vec{A}_\perp(\tau,\vec{\sigma}')
\right]^s\,{\bf P}^{sr}(\vec{\sigma}',\vec{\sigma}),\nonumber\\
&&\nonumber\\
\frac{\partial \pi^r_\perp(\tau,\vec{\sigma})}{\partial\tau}&=&
\Delta {\tilde A}^r_\perp(\tau,\vec{\sigma}) -
 {1\over c}\, \int d^3\sigma'\, \left[ - \vec{\Omega} \cdot \vec{\sigma}'
 \times {\vec \partial}'\, \vec{\pi}_\perp(\tau,\vec{\sigma}') + \vec{\Omega}
\times \vec{\pi}_\perp(\tau,\vec{\sigma}')
\right]^s\, {\bf P}^{sr}(\vec{\sigma}',\vec{\sigma}) +\nonumber\\
&&\nonumber\\
&+&\sum_i\, Q_i\, \left(\dot{\vec{\eta}}_i(\tau ) + \vec{\Omega}
\times \vec{\eta}_i(\tau )\right)^s\, {\bf
P}^{sr}(\vec{\eta}_i,\vec{\sigma}).
 \label{6.1}
 \eea

\hfill

For the study of homogeneous solutions of these equations, i.e. for
incoming electro-magnetic waves propagating in regions where there
are no charged particles, these equations can be replaced with the
following ones (we use the vector notation of  Section IVC of paper
I)

 \bea
\frac{\partial \vec{A}_\perp(\tau,\vec{\sigma})}{\partial\tau}&=&
\vec{\pi}_\perp(\tau,\vec{\sigma}) -
  {1\over c}\, \left[ - \vec{\Omega} \cdot \vec{\sigma} \times \vec
\partial\, \vec{A}_\perp(\tau,\vec{\sigma}) + \vec{\Omega}
\times \vec{A}_\perp(\tau,\vec{\sigma})\right],\nonumber\\
&&\nonumber\\
\frac{\partial \vec{\pi}_\perp(\tau,\vec{\sigma})}{\partial\tau}&=&
\Delta \vec{A}_\perp(\tau,\vec{\sigma}) -
 {1\over c}\, \left[ - \vec{\Omega} \cdot \vec{\sigma} \times \vec
\partial\, \vec{\pi}_\perp(\tau,\vec{\sigma}) + \vec{\Omega} \times
\vec{\pi}_\perp(\tau,\vec{\sigma})\right].
 \label{6.2}
 \eea

\hfill

As shown in Appendix A of I, this result allows to recover the form
given by Schiff in Appendix A of ref.\cite{28} for the
Landau-Lifschitz non-inertial electro-magnetic fields \cite{17}.
\bigskip

Let us look at solutions of Eqs.(\ref{6.2}) in the following two
ways.

\subsubsection{Going back to an Inertial Frame}

Let us look at solution by reverting to an inertial frame.\medskip

By introducing the 3-coordinates

 \beq
 X^a(\tau) = R^a{}_r(\tau)\, \sigma^r,
  \label{6.3}
 \eeq

\noindent at each value of $\tau$ by means of a $\tau$-dependent
rotation (it would become also point-dependent if we go beyond rigid
rotations) we can go from the rigidly rotating non-inertial frame
with radar 4-coordinates $(\tau ; \sigma^u)$ to an instantaneously
comoving inertial frame, centered on the same inertial observer,
with 4-coordinates $(\tau; X^a)$.
\medskip

Let us assume that the non-inertial transverse electromagnetic
potential $A_{\perp\, r}(\tau ,\sigma^u)$ can be obtained from the
instantaneously comoving inertial transverse potential
$A^{(com)}_{\perp\, a}(\tau , X^a(\tau))$ by using the rotation
matrix $R(\tau)$

 \beq
 A_{\perp\, r}(\tau, \sigma^u) = A^{(com)}_{\perp\, a}\Big(\tau, X^a(\tau) =
 R^a{}_s(\tau)\, \sigma^s\Big)\, R^a{}_r(\tau).
 \label{6.4}
  \eeq

\noindent By definition $A^{(com)}_{\perp\, a}(\tau , X^a(\tau))$
satisfies the inertial Maxwell equations in the radiation gauge
(obtainable by putting Eqs.(\ref{6.4}) into Eqs.(\ref{6.2}))

 \beq
  \frac{\partial^2\, A^{(com)}_{\perp\, a}(\tau, X^b)}{\partial\tau^2} -
  \Delta_X\, A^{(com)}_{\perp\, a}(\tau, X^b) = 0,\qquad
\sum_a\,\frac{\partial}{\partial X^a}\, A^{(com)}_{\perp\, a}(\tau,
X^b) = 0.
 \label{6.5}
  \eeq

\bigskip

This result is in accord with the general covariance of non-inertial
Maxwell equations and is also consistent with the locality
hypothesis (see  Section IIB of paper I) of the the 1+3 approach.

\bigskip
If we consider the following plane wave solution with constant $F_a$
and ${\hat K}_a$ and $\sum_a\, {\hat K}_a\, F_a = 0$ (transversality
condition)

 \beq
 A^{(com)}_{\perp\, a}(\tau, X^b) = \frac{1}{\omega}\, F_a\,
 e^{i\, \frac{\omega}{c}\, \left(\tau - \sum_a\, {\hat{K}}_a\, X^a\right)},
 \label{6.6}
  \eeq

\noindent we get the following expression for the non-inertial
solution

   \bea
 A_{\perp\, r}(\tau, \sigma^u) &=& F_a\, R^a{}_r(\tau)\, e^{
i\, \frac{\omega}{c}\, \Phi(\tau, \sigma^u)},\nonumber \\
 &&{}\nonumber \\
 \Phi(\tau, \sigma^u) &=& \tau - \hat{K}_a\, R^a{}_r(\tau)\, \sigma^r
 \approx{|}_{\vec \Omega = const.}\, \tau\, \Big(1 + {{\vec \Omega}\over c}\,
 \cdot \vec \sigma \times \hat K\Big) -
 {\hat K} \cdot \vec \sigma + O(\Omega^2/c^2).
 \label{6.7}
 \eea

\subsubsection{Eikonal Approximation}

Let us now look at solutions  by making the following eikonal
approximation (without any commitment with the locality hypothesis)

 \beq
  A_{\perp\, r}(\tau, \sigma^u) =
\frac{1}{\omega}\, a_r(\tau, \sigma^u)\, e^{i\, {{\omega}\over c}\,
\Phi(\tau, \sigma^u)} + O(1/\omega^2).
 \label{6.8}
  \eeq

\noindent and by putting this expression in Eqs.(\ref{6.2}).
\medskip

Let us consider the case in which we have $\omega/c
>> 1$ e $\Omega/c << 1$, so that Eqs.(\ref{6.2}) become a power
series in $\omega/c$. By neglecting terms in $\Omega^2/c^2$ and
terms in $(c/\omega)^{-k}$ for $k \geq 0$, the dominant terms are:

a) at the order $\omega/c$ the equation for the phase $\Phi$, named
{\em eikonal equation};

b) at the order $(\omega/c)^o=1$ the equation for the amplitude
$a_r$, named {\em first-order transport equation}.
\medskip

These equations have the following form ($\vec a = \{ a_r \}$)

 \bea
 &&\Big[\left(\frac{\partial\Phi}{\partial\tau}\right)^2 - 2\,{{\vec{\Omega}}\over c} \cdot
 \vec{\sigma} \times \vec \partial\, \Phi - \left(\vec
 \partial\, \Phi\right)^2\Big](\tau ,\sigma^u) + O(\Omega^2/c^2)
 = 0\nonumber\\
&&\nonumber\\
&&\nonumber\\
&&\Big[\frac{\partial\Phi}{\partial\tau}\,\left(\frac{\partial\vec{a}}{\partial\tau}
+ {{\vec{\Omega}}\over c} \times \vec{a} - {{\vec{\Omega}}\over c}
\cdot \vec{\sigma} \times\vec{\partial} \,\vec{a}\right) -
\frac{\partial\vec{a}}{\partial\tau}\, {{\vec{\Omega}}\over c} \cdot
\vec{\sigma} \times \vec
\partial\, \Phi - \left(\vec \partial\, \Phi \cdot \vec
\partial \right)\, \vec{a}\Big](\tau ,\sigma^u) =\nonumber\\
\nonumber\\
&=&- \frac{1}{2}\, \left(\frac{\partial^2\Phi}{\partial\tau^2} -
2\,(\vec{\Omega} \times\vec{\sigma} \cdot \vec{\partial})\,
\frac{\partial\Phi}{\partial\tau} - \triangle\Phi\right)(\tau
,\sigma^u) + O(\Omega^2/c^2)
\nonumber\\
&&\nonumber\\
&&\nonumber\\
&& \Big[\vec a \cdot \vec \partial\, \Phi\Big](\tau ,\sigma^u) =
0\quad (transversality\, condition).
 \label{6.9}
 \eea

Let us look for solutions of the  {\em eikonal equation} for $\Phi$
of the form

 \beq
\Phi(\tau, \sigma^u) = \tau + F(\sigma^u),
 \label{6.10}
  \eeq

\noindent where we have chosen the boundary condition

 \beq
  \frac{\partial\Phi}{\partial\tau}=1.
   \label{6.11}
 \eeq

This condition implies that the solution of Eq.(\ref{6.8}) describes
a ray emitted from a source having a characteristic frequency
$\omega$ when it is at rest in the non-inertial frame. Let us remark
that in more general cases this type of boundary conditions are
possible only if the {\em 3-metric} $h_{rs}$ and the  {\em lapse}
($n$) and {\em shift} ($n^r$) functions are stationary in the
non-inertial frame.\medskip

 An expansion in powers of $\Omega/c$ of $F(\sigma^u)$, namely
$F(\sigma^u) = F_o(\sigma^u) + \frac{\Omega}{c}\, F_1(\sigma^u) +
O\Big(\frac{\Omega^2}{c^2}\Big)$, gives the following form of the
eikonal equation

 \beq
\Big[1 - \Big(\vec{\partial}\, F_o(\sigma^u)\Big)^2\Big] -
\frac{2\Omega}{c}\, \Big[ \hat{\Omega} \cdot \vec{\sigma}
\times\vec{\partial}F_o(\sigma^u) + \vec{\partial}\, F_o(\sigma^u)
\cdot \vec{\partial}\, F_1(\sigma^u) \Big] +
O\Big(\frac{\Omega^2}{c^2}\Big) = 0,
 \label{6.12}
 \eeq

\noindent implying:

a) the equation $1 - \Big(\vec{\partial}\, F_o(\sigma^u)\Big)^2 = 0$
at the order zero in $\Omega$. If $\hat k$ is an arbitrary unit
vector (the propagation direction of the plane wave in the inertial
limit $\Omega \mapsto 0$), its solution is

 \beq
 F_o(\sigma^u) = - \hat{k} \cdot \vec{\sigma}.
 \label{6.13}
 \eeq

\medskip

b) the equation $\hat{k} \cdot \vec{\partial}\, F_1(\sigma^u) = -
\hat{\Omega} \cdot \vec{\sigma} \times \hat{k}$ for $F_1(\sigma^u)$,
after having used Eq.(\ref{6.13}), at the order one in $\Omega$.
Since we have $(\hat{k} \cdot \vec{\partial})\, (\hat{\Omega} \cdot
\vec{\sigma} \times \hat{k}) = 0$ and $(\hat{k} \cdot
\vec{\partial})\, (\hat{k} \cdot \vec{\sigma}) = 1$, the solution
for $F_1(\sigma^u)$ is

 \beq
  F_1(\sigma^u) = - \Big(\hat{\Omega}
\cdot \vec{\sigma} \times \hat{k}\Big)\, (\hat{k} \cdot
\vec{\sigma}).
 \label{6.14}
 \eeq

\medskip

Therefore the solution for $\Phi$ is

 \beq
\Phi(\tau,\sigma^u) = \tau - \hat{k} \cdot \vec{\sigma}\, \Big(1 +
\vec{\Omega} \cdot\vec{\sigma} \times \hat{k}\Big).
 \label{6.15}
  \eeq

\bigskip

The phases in the solutions (\ref{6.7}) and (\ref{6.15}) of
Eqs.(\ref{6.2}) are different since the solutions have different
boundary conditions. The solution (\ref{6.7}) satisfies also the
eikonal equation but not the boundary condition (\ref{6.11}), since
we have ${{\partial\, \Phi}\over {\partial\, \tau}} = 1 - {\hat
K}_a\, R^a{}_r(\tau )\, \epsilon_{ruv}\, {\tilde \Omega}^u\,
\sigma^v \not= 1$.

\bigskip

Let us remark that both the solutions (\ref{6.7}) and (\ref{6.15})
have the following structure

 \beq
\tilde{A}^r_\perp(\tau, \sigma^u) \sim\, {\cal A}^r(\tau,
\sigma^u)\, e^{i\, \varphi(\tau, \sigma^u)},
 \label{6.16}
  \eeq

\noindent where ${\cal A}^r(\tau, \sigma^u) \sim\,  O(1/\omega)$ is
the {\em amplitude} and $\varphi(\tau, \sigma^u) \sim  O(\omega)$ is
the {\em phase}. The only difference is that the solution
(\ref{6.7}) holds for every value of $\omega$ (also for the small
values corresponding to the radio waves of the GPS system), while
the solution (\ref{6.15}) for the phase of the eikonal approximation
(\ref{6.8}) holds only for higher values of $\omega$, corresponding
to visible light.

\subsubsection{Light Rays}

Given the phase of Eq.(\ref{6.16}), the trajectories of the light
rays are defined as the lines orthogonal (with respect to the
4-metric $g_{AB}$ of the 3+1 splitting) to the hyper-surfaces
$\varphi(\tau, \sigma^u) = \,const.$. Therefore the trajectories
$\sigma^A(s)$ ($s$ is n affine parameter) satisfy the equation

 \beq
\frac{d\sigma^A(s)}{ds} = g^{AB}(\sigma(s))\, \,
\frac{\partial\varphi}{\partial\sigma^B}(\sigma(s)).
 \label{6.17}
 \eeq

For instance in the case of our rigidly rotating foliation, for
which Eqs.(2.14)-(2.16) of I imply $g^{\tau\tau} = 1$, $g^{\tau r} =
- (\vec{\Omega} \times \vec{\sigma})^r$, $g^{rs} = - \delta^{rs} +
O(\Omega^2/c^2)$, Eqs.(\ref{6.17}) take the form

 \bea
\frac{d\tau(s)}{ds}&=&\omega\, + \vec{k} \cdot
\left(\frac{\vec{\Omega}}{c}
\times \vec{\sigma}\right) + O(\Omega^2/c^2),\nonumber\\
&&\nonumber\\
\frac{d\sigma^r(s)}{ds}&=&\omega\, \left(\frac{\vec{\Omega}}{c}
\times \vec{\sigma}\right)^r + k^r\, \left(1 +
\frac{\vec{\Omega}}{c} \times \vec{\sigma} \cdot \hat{k}\right) -
\left(\frac{\vec{\Omega}}{c} \times \hat{k}\right)^r\, (\vec{k}
\cdot \vec{\sigma}) + O(\Omega^2/c^2),\nonumber \\
 &&{}
 \label{6.18}
 \eea

\noindent whose solution has the  form

 \beq
\vec{\sigma}(\tau) - \vec{\sigma}(0) = \hat{k}\, \tau +
\left(\frac{\vec{\Omega}}{c} \times \hat{k}\right)\, \tau^2 +
O(\Omega^2/c^2).
 \label{6.19}
  \eeq

This equation shows that in the rotating frame the ray of light
appears to deviate from the {\it inertial} trajectory
$\vec{\sigma}(\tau) = \hat{k}\, \tau$ due to the {\it centrifugal
correction} $\vec{c}(\tau) = \left(\frac{\vec{\Omega}}{c} \times
\hat{k}\right)\, \tau^2 + O(\Omega^2/c^2)$ implying $\hat{k} \cdot
\vec{c}(\tau) = 0 + O(\Omega^2/c^2)$.

\subsection{Sources and Detectors}

To connect the previous solutions to the interpretation of observed
data we need a schematic description of {\it sources} and {\it
detectors}.

In many applications {\em sources} and {\em detectors} are described
as point-like objects, which follow a prescribed world-line $
\zeta^A(\tau) = (\tau, \eta^u(\tau))$ with unit 4-velocity
$v^A(\tau) = \frac{d \zeta^A(\tau)}{d\tau}\,
\left(g_{CD}(\zeta(\tau))\, \frac{d \zeta^C(\tau)}{d\tau}\, \frac{d
\zeta^D(\tau)}{d\tau}\right)^{-1/2} $.

This description is enough for studying the influence of the
relative motion between source and detector on the frequency emitted
from the source and that observed by the detector (it works equally
well for the Doppler effect and for the gravitational redshift in
presence of gravity). With solutions like Eq.(\ref{6.16}) the
frequency emitted by a source located in $\zeta_s{}^A$ and moving
with 4-velocity  $v_s{}^A$ and that observed by a detector in
$\zeta_r{}^A$ and moving with 4-velocity  $v_r{}^A$ are $\omega_s =
v_s{}^A\, \partial_A\, \varphi(\zeta_s)$ and $\omega_r = v_r{}^A\,
\partial_A\, \varphi(\zeta_r)$, respectively.

This justifies the boundary condition (\ref{6.11}), because sources
at rest in the rotating frame with coordinates $(\tau,\sigma^r)$
have 4-velocity $v^A = (1,0)$.

\medskip

However, to measure the electro-magnetic field in assigned (spatial)
polarization direction we must assume that the detector is endowed
with a tetrad orthonormal with respect to the 4-metric of the 3+1
splitting, such that the time-like 4-vector is the unit 4-velocity
of the detector: in 4-coordinates adapted to the 3+1 splitting they
are ${\cal E}^A_{(\alpha)}(\tau) = \Big({\cal E}^A_{(o)}(\tau ) =
v^A(\tau); {\cal E}^A_{(i)}(\tau)\Big)$, $g_{AB}(\zeta_r(\tau))\,
{\cal E}^A_{(\alpha)}(\tau)\, {\cal E}^B_{(\beta)}(\tau) =
\eta_{(\alpha)(\beta)}$ (see Section IIB of paper I for the 1+3
point of view). A detector measures the following field strengths
along the spatial polarization directions ${\cal E}^A_{(i)}(\tau)$:
$\check{E}_{(i)} = F_{AB}\, v^A\, {\cal E}^B_{(i)}$ and
$\check{B}_{(i)} = (1/2)\, \epsilon_{(i)(j)(k)}\, F_{AB}\, {\cal
E}^A_{(j)}\, {\cal E}^B_{(k)}$.

\medskip

Let us consider the following two cases.

\subsubsection{Detectors at Rest in an Inertial Frame}

A detector at rest in the instantaneous inertial frame with
coordinates $(\tau; X^a(\tau))$ follows the straight world-line
$\zeta^{\mu}_{r,in}(\tau) = \tau\, \epsilon^{\mu}_{\tau} +
\epsilon^{\mu}_a\, \eta_{in}^a$ with $\eta_{in}^a = const.$ and has
the 4-velocity $u^{\mu} = \epsilon^{\mu}_{\tau}$. If the reference
asymptotic tetrad $\epsilon^{\mu}_A$ of the foliation is related by
$\epsilon^{\mu}_A = \Lambda^{\mu}_{(o)}{}_{\nu}\, e^{\nu}_{(A)}$ to
a tetrad $e^{\mu}_{(A)} = \delta^{\mu}_A$  aligned to the axes of
the inertial frame in Cartesian coordinates, then a generic
time-independent non-rotating tetrad associated with the detector
will be ${\cal G}^{\mu}_{(A)} = \Lambda_{(A)}{}^{(B)}\,
e^{\mu}_{(B)} = \Lambda_{(A)}{}^{(\mu)}$ if ${\cal G}^{\mu}_{(\tau)}
= u^{\mu}$. Here the $\Lambda$'s denote Lorentz transformations. The
detector will measure the standard electric and magnetic fields
$\check{E}_{(i)} = F_{\mu\nu}\, u^{\mu}\, {\cal G}^{\nu}_{(i)}$ and
$\check{B}_{(i)} = (1/2)\, \epsilon_{(i)(j)(k)}\, F_{\mu\nu}\, {\cal
G}^{\mu}_{(j)}\, {\cal G}^{\nu}_{(k)}$.

\subsubsection{Sources and Detectors at Rest in Rotating Frames}

Lt us now consider sources and detectors at rest in the nearly rigid
rotating frame described by the embedding $z^{\mu}(\tau ,\sigma^u) =
\epsilon^{\mu}_{\tau}\, \tau + \epsilon^{\mu}_r\, R^r{}_s(\tau)\,
\sigma^s + O(\Omega^2/c^2)$, so that $z^{\mu}_{\tau}(\tau ,\sigma^u)
= \epsilon^{\mu}_{\tau} + \epsilon^{\mu}_r\, {\dot R}^r{}_s(\tau)\,
\sigma^s + O(\Omega^2/c^2)$ and $z^{\mu}_r(\tau ,\sigma^u) =
\epsilon^{\mu}_s\, R^s{}_r(\tau) + O(\Omega^2/c^2)$.
\medskip

The world-line of these objects will have the form $\zeta^{\mu}(\tau
) = \tau\, \epsilon^{\mu}_{\tau} + \epsilon^{\mu}_r\, R^r{}_s(\tau
)\, \eta_o^s + O(\Omega^2/c^2) = \epsilon^{\mu}_A\, \zeta^A(\tau)$
with $\eta^r_o = const.$. We have $\zeta^{\tau}(\tau) = \tau$ and
$\zeta^r(\tau) = R^r{}_s(\tau )\, \eta_o^s + O(\Omega^2/c^2)$.
Therefore these objects coincide with some of the observers
belonging at the non-surface forming congruence generated by the
evolution vector field as said in  Section IIB of paper I. Since the
world-lines of the Eulerian observers of the other congruence are
not explicitly known, it is not possible to study the behavior of
objects coinciding with some of these observers.
\medskip

Therefore the unit 4-velocity $u^{\mu}(\tau) = \epsilon^{\mu}_A\,
v^A(\tau)$ will have the components $v^A(\tau)$  proportional to
${\dot \zeta}^A(\tau) = \Big(1; {\dot R}^r{}_s(\tau )\, \eta_o^s +
O(\Omega^2/c^2)\Big) \approx{|}_{\vec \Omega = const.}\,\, \Big(1;
R^r{}_s(\tau)\, ({\vec \eta}_o \times \vec {{\Omega}\over c})^s
+O(\Omega^2/c^2)\Big)$, where the definitions after Eq.(2.14) of I
have been used.\medskip

We can also write $u^{\mu}(\tau) = {\tilde u}^A(\tau)\,
z^{\mu}_A(\tau, \eta_o^u)$ by using the non-orthonormal tetrads
$z^{\mu}_A(\tau ,\sigma^u)$. Then we get  $v^{\tau}(\tau) = {\tilde
u}^{\tau}(\tau) + O(\Omega^2/c^2)$ and $v^r(\tau) = {\tilde
u}^{\tau}(\tau)\, {\dot R}^r{}_s(\tau)\, \eta_o^s + R^r{}_s(\tau)\,
{\tilde u}^s(\tau) + O(\Omega^2/c^2)$. While the quantities
$v^A(\tau)$ give the description of the 4-velocity with respect to
the asymptotic non-rotating inertial observers, the quantities
${\tilde u}^A(\tau)$ explicitly show the effect of the rotation at
the position $\eta_o^r$ of the object. Therefore it should be
${\tilde u}^A(\tau) = (1; 0)$ at the lowest order: indeed we get
${\tilde u}^{\tau}(\tau) = 1 + O(\Omega^2/c^2)$ and ${\tilde
u}^r(\tau) = v^s(\tau)\, R_s{}^r(\tau) - {\tilde u}^{\tau}\,
\Big(R^{-1}(\tau)\, {\dot R}(\tau)\Big)^r{}_s\, \eta^s_o   = 0 +
O(\Omega^2/c^2)$.\medskip

For the constant unit normal to the instantaneous 3-spaces we get
$l^{\mu} = \epsilon^{\mu}_{\tau} = {\tilde l}^A(\tau, \eta_o^r)\,
z^{\mu}(\tau, \eta_o^r)$ with ${\tilde l}^{\tau}(\tau ,\eta_o^r) = 1
+ O(\Omega^2/c^2)$ and ${\tilde l}^r(\tau, \eta_o^u) = - {\tilde
l}^{\tau}(\tau ,\eta_o^u)\, \Big(R^{-1}(\tau)\, {\dot
R}(\tau)\Big)^r{}_s\, \eta^s_o = - ({\vec \eta}_o \times {{\vec
\Omega}\over c})^r + O(\Omega^2/c^2)$.\medskip

Let us introduce an orthonormal tetrad ${\cal W}^{\mu}_{(\alpha)}$,
$\eta_{\mu\nu}\, {\cal W}^{\mu}_{(\alpha)}\, {\cal
W}^{\nu}_{(\beta)} = \eta_{(\alpha)(\beta)}$, whose time-like
4-vector is $l^{\mu}$, i.e. We have ${\cal W}^{\mu}_{(o)} = l^{\mu}
= \epsilon^{\mu}_{\tau} = {\cal W}^A_{(o)}\, \epsilon^{\mu}_A =
{\tilde {\cal W}}^A_{(o)}(\tau ,\eta_o^u)\, z^{\mu}_A(\tau
,\eta_o^u)$ with .${\cal W}^A_{(o)} = (1; 0)$ and ${\tilde {\cal
W}}^A_{(o)}(\tau ,\eta_o^u) = {\tilde l}^A(\tau ,\eta_o^u) = \Big(1;
- ({\vec \eta}_o \times {{\vec \Omega}\over c})^r\Big) +
O(\Omega^2/c^2)$. The spatial axes ${\cal W}^{\mu}_{(i)} = {\cal
W}^A_{(i)}\, \epsilon^{\mu}_A = {\tilde {\cal W}}^A_{(i)}(\tau
,\eta_o^u)\, z^{\mu}_A(\tau ,\eta_o^u)$ with $l_{\mu}\, {\cal
W}^{\mu}_{(i)} = [{\tilde l}^A\, g_{AB}\, {\tilde {\cal W}}^B](\tau
,\eta_o^u) = 0$ must be non-rotating with respect to the observer
with 4-velocity proportional to $z^{\mu}_{\tau}(\tau ,\eta_o^u)$.
Therefore we must have ${\tilde {\cal W}}^A_{(i)} = \Big(0; {\tilde
{\cal W}}^r_{(i)}\Big)$ with ${\tilde {\cal W}}^r_{(i)} = const.$.
As a consequence we have ${\cal W}^A_{(i)}(\tau) = \Big(0;
R^r{}_s(\tau)\, {\tilde {\cal W}}^s_{(i)} \Big) + O(\Omega^2/c^2)$.

\medskip

The polarization axes of sources and detectors will be defined by a
tetrad ${\cal E}^{\mu}_{(\alpha)}(\tau ,\eta_o^r) = {\cal
E}^A_{(\alpha)}(\tau ,\eta_o^r)\, \epsilon^{\mu}_A = {\tilde {\cal
E}}^A_{(\alpha)}(\tau ,\eta_o^r)\, z^{\mu}_A(\tau ,\eta_o^r)$,
$\eta_{\mu\nu}\, {\cal E}^{\mu}_{(\alpha)}\, {\cal
E}^{\nu}_{(\beta)} = \eta_{(\alpha)(\beta)}$ with the following
properties:\medskip

a) the time-like 4-vector ${\cal E}^{\mu}_{(o)}(\tau ,\eta_o^r)$ is
such that its components ${\tilde {\cal E}}^A_{(o)}(\tau ,\eta_o^r)$
coincide with the components ${\tilde u}^A(\tau) = (1; 0) +
O(\Omega^2/c^2)$ of the 4-velocity $u^{\mu}(\tau)$ of the object
located at $\zeta^{\mu}(\tau) = z^{\mu}(\tau ,\eta_o^r)$: as a
consequence we have ${\cal E}^{\mu}_{(o)}(\tau ,\eta_o^r) =
z^{\mu}_{\tau}(\tau ,\eta_o^r) + O(\Omega^2/c^2) = u^{\mu}(\tau)$;

b) the spatial axes ${\cal E}^{\mu}_{(i)}(\tau ,\eta_o^r)$,
orthogonal to the 4-velocity $u^{\mu}(\tau)$, must be at rest in the
rotating frame: we have to identify their components ${\tilde {\cal
E}}^A_{(i)}(\tau ,\eta_o^r)$.
\medskip

If at the observer position we consider the Lorentz transformation
sending $l^{\mu}$ to $u^{\mu}(\tau )$, i.e. $L^{\mu}{}_{\nu}(l
\mapsto u(\tau))$, its projection  $L^A{}_B(\vec \beta) \,
{\buildrel {def}\over =}\, \epsilon^A_{\mu}\, L^{\mu}{}_{\nu}(l
\mapsto u(\tau))\, \epsilon^{\nu}_B$ is a Wigner boost, see Eq.(2.8)
of I, with parameter $\vec \beta = \{ \beta^r = R^r{}_s(\tau )\,
\Big({\vec \eta}_o \times {{\vec \Omega}\over c}\Big)^s$ (so that
$\gamma = \sqrt{1 - {\vec \beta}^2} = 1 + O(\Omega^2/c^2)$).
Therefore the transformation sending the components ${\tilde
l}^A(\tau ,\eta_o^u)$ of the unit normal into the components
${\tilde u}^A(\tau)$ of the 4-velocity modulo terms of order
$O(\Omega^2/c^2)$ is

\bea
 {\tilde {\cal E}}^A_{(o)}(\tau, \eta_o^u) &=& {\tilde u}^A(\tau) =
(1; 0) + O(\Omega^2/c^2) =\nonumber \\
 &=&\Big(z^A_{\mu}\, \epsilon^{\mu}_C\, L^C{}_D(\vec \beta)\,
\epsilon_{\nu}^D\, z^{\nu}_B\, {\tilde l}^B_{(o)}\Big)(\tau
,\eta_o^u) + O(\Omega^2/c^2) =\nonumber \\
 &=&\Big(z^A_{\mu}\, \epsilon^{\mu}_C\, L^C{}_D(\vec \beta)\,
 \epsilon_{\nu}^D\, z^{\nu}_B\, {\tilde {\cal W}}^B_{(o)}\Big)(\tau ,\eta_o^u)
 + O(\Omega^2/c^2),\nonumber \\
 &&{}\nonumber \\
 {\tilde {\cal E}}^A_{(i)}(\tau, \eta_o^u) &=&\Big(z^A_{\mu}\, \epsilon^{\mu}_C\,
 L^C{}_D(\vec \beta)\, \epsilon_{\nu}^D\, z^{\nu}_B\,
 \Big)(\tau ,\eta_o^u)\, {\tilde {\cal
W}}^B_{(i)}.
 \label{6.20}
 \eea

This complete the construction of the non-rotating tetrads ${\cal
E}^{\mu}_{(\alpha)}(\tau ,\eta_o^u)$ for the objects at rest at
$\eta_o^r$.\medskip

A detector endowed of such a non-rotating tetrad will measure the
following projections of the electro-magnetic field strength on its
polarization directions

 \beq
\hat{E}_{(i)} = F_{AB}\, {\tilde u}^A\, {\tilde {\cal
E}}^B_{(i)},\qquad \hat{B}_{(i)} = {1\over 2}\,
\epsilon_{(i)(j)(k)}\, F_{AB}\, {\tilde {\cal E}}^A_{(j)}\, {\tilde
{\cal E}}^B_{(k)}.
 \label{6.21}
 \eeq

These quantities have to be confronted with the non-inertial
electric and magnetic fields $E_r$ and $B_r$, whose projections on
the non-rotating spatial axes ${\tilde {\cal W}}^A_{(i)} = (0;
{\tilde {\cal W}}^r_{(i)})$ inside the instantaneous 3-space are

 \beq
{E}_{(i)} = E_r\, {\tilde {\cal W}}^r_{(i)},\qquad {B}_{(i)} = B_r\,
{\tilde {\cal W}}^r_{(i)}.
 \label{6.22}
  \eeq

Eqs.(\ref{6.20}) imply the following connection among these
quantities

 \bea
  \hat{E}_{(i)}&=& {E}_{(i)} + {\cal
O}(\Omega^2/c^2),\nonumber\\
&&\nonumber\\
\hat{B}_{(i)}&=& {B}_{(i)} - \epsilon_{ijk}\, {\tilde {\cal
W}}^r_{(j)}\, \delta_{rs}\, \Big({\vec \eta}_o \times {{\vec
\Omega}\over c}\Big)^s\, {E}_{(k)} + {\cal O}(\Omega^2/c^2).
 \label{6.23}
 \eea

\medskip

For radio wave (like in the case of GPS) the directions ${\cal
G}_{(i)}^a$ or ${\tilde {\cal E}}_{(i)}^r$ are realized by means of
antennas attached to both emitters and receivers. In the optical
range the antennas are replaced by components of the macroscopic
devices used for the emission and the detection.

\subsection{The Phase Wrap Up Effect}

The  {\em phase wrap up} is a  modification of the phase when a
receiver in rotational motion analyzes the circularly polarized
radiation emitted by a source at rest in an inertial frame. Till now
the effect has been explained by using the 1+3 point of view and the
locality hypothesis in Refs.\cite{31}, where it shown that it is a
particular case of helicity-rotation coupling (the spin-rotation
coupling for photons). It has been verified experimentally, in
particular in GPS \cite{36}, where the receiving antenna on the
Earth surface is rotating with Earth.

\medskip

We will explain the effect by using the non-inertial solution
(\ref{6.7}) and an observer at rest in an inertial frame endowed of
the tetrad ${\cal G}^{\mu}_{(A)}$ defined in Subsubsection 1 of
Subsection B. We rewrite the spatial axes in the form $ {\cal
G}^a_{(i)} = \Big(I^a_{(1)}, I^a_{(2)}, \hat{K}^a\Big)$ with the
vectors satisfying $\vec{I}_{(1)} \cdot \vec{I}_{(2)} = 0$,
$\vec{I}_{(\lambda)} \cdot \hat{K} = 0$ ($\lambda = 1,2$),
$\vec{I}_{(\lambda)}^{\,2} = 1$. Then we pass to a circular basis by
introducing the vectors $\vec{I}_{(\pm)} = \frac{\vec{I}_{(1)} + i\,
\vec{I}_2}{\sqrt{2}}$, which satisfy $\hat{K} \cdot \vec{I}_{(\pm)}
= 0$, $\vec{I}_{(\pm)}^{\,2} = 0$ and $\vec{I}_{(+)} \cdot
\vec{I}_{(-)} = 1$.\medskip

In the rotating non-inertial frame a right-circularly polarized
wave, emitted in the inertial frame, will have the form (\ref{6.7})
($\hat{K} \cdot \vec{I}_{(+)} = 0$ is the transversality condition)

\beq A_{\perp r}(\tau,\vec{\sigma}) = \frac{F}{\omega}\, I_{(+)a}\,
R^a{}_r(\tau)\, e^{i\, {{\omega}\over c}\, \Phi(\tau,\vec{\sigma})}.
 \label{6.24}
 \eeq
\medskip

Let us remark that in the circular basis we have ${\vec A}_{\perp} =
A_n\, \hat n + A_+\, {\vec I}_{(+)} + A_{-}\, {\vec I}_{(-)}$, but
the components $A_n$, $A_{\pm}$, coincide with either linearly or
circularly polarized states of the electro-magnetic field only for
$\hat{n} = \hat{k}$, since $\hat K = {{\omega}\over c}\, \hat k$
(${\hat K}^2 = {{\omega^2}\over {c^2}}$) is the wave vector.

From Eqs(\ref{6.24}) we obtain the following non-inertial magnetic
and electric fields (2.19) of I

 \bea
B_r&=&-\, \frac{F}{c}\, I_{(+)a}\, R^a{}_r(\tau)\, e^{i\,
{{\omega}\over c}\, \Phi(\tau,\vec{\sigma})}\byd B_{o}\,
I_{(+)a}(K)\, R^a{}_r(\tau)\, e^{i\, {{\omega}\over
c}\, \Phi(\tau,\vec{\sigma})},\nonumber\\
&&\nonumber\\
E_r&=&- i\, \frac{F}{c}\,\, I_{(+)a}\, R^a{}_r(\tau)\, e^{i\,
{{\omega}\over c}\, \Phi(\tau,\vec{\sigma})} + \frac{1}{c}\,
(\vec{\Omega}\times\vec{\sigma}) \times \vec{B} =\nonumber \\
 &{\buildrel {def}\over =}& E_{o}\, I_{(+)a}\, R^a{}_r(\tau)\, e^{i\,
 {{\omega}\over c}\, \Phi(\tau,\vec{\sigma})}
 +\, E_\ell\, \hat{K}_a\, R^a{}_r(\tau)\,\, e^{i\,
{{\omega}\over c}\, \Phi(\tau,\vec{\sigma})},\nonumber \\
 &&{}\nonumber \\
  B_o = - \frac{F}{c},&& E_o = - i\, \frac{F}{c} + \frac{1}{c}\,
(\vec{\Omega}\times\vec{\sigma}) \times \vec{B} \cdot \vec{I}_{(-)},
\quad E_\ell = \frac{1}{c}\, (\vec{\Omega} \times \vec{\sigma})
\times \vec{B} \cdot \hat{K}.
 \label{6.25}
 \eea

\bigskip

Let us now consider a receiver at rest in the rotating frame. Since
its 4-velocity is ${\tilde u}^A = (1; 0)$, it can be endowed with
the non-rotating tetrad ${\tilde {\cal W}}^A_{(\alpha)}$ of
Subsubsection 2 of Subsection B. If $\hat n$ is the unit vector in
the direction of the rotation axis, i.e. if $\vec \Omega = \Omega\,
\hat n$, we can choose the spatial axes ${\tilde {\cal W}}^r_{(i)} =
(\epsilon^r_{(1)}, \epsilon^r_{(2)}, \hat{n}^r)$ with
$\vec{\epsilon}_{(1)} \cdot \vec{\epsilon}_{(2)} = 0$,
$\vec{\epsilon}_{(\lambda)} \cdot \hat{K} = 0$,
$\vec{\epsilon}_{(\lambda)}^{\,2} = 1$. If we introduce the circular
basis $\vec{\epsilon}_{(\pm)} = \frac{\vec{\epsilon}_{(1)} + i\,
\vec{\epsilon}_2}{\sqrt{2}}$, we have $\hat{n} \cdot
\vec{\epsilon}_{(\pm)} = 0$, $\vec{\epsilon}_{(\pm)}^{\,2} = 0$,
$\vec{\epsilon}_{(+)} \cdot \vec{\epsilon}_{(-)} = 1$ and
$R^a{}_r(\tau)\, \epsilon_{(\pm)}{}^r = \epsilon_{(\pm)}^a\,\,
e^{\left[\pm\, i\, {{\Omega}\over c}\, \tau\right]}$.
\bigskip

The receiver will measure the following magnetic and electric fields

 \begin{eqnarray*}
B_n&=&B_r\,\hat{n^r}=B_o
\,\left(\vec{I}_{(+)a} \hat{n}^a\right)\, \exp\,\left[i {{\omega}\over c}\,\Phi\right],\nonumber\\
&&\nonumber\\
B_{(\pm)}&=&B_r\,\epsilon^r_{(\mp)}=B_o \, \left(\vec{I}_{(+)a}
{\epsilon}^a_{(\mp)}\right)\,\exp\,\left[{i\over c}\, \left(\mp
\Omega\, \tau + \omega\, \Phi(\tau,\vec{\sigma})\right)\right],
 \end{eqnarray*}

 \bea
E_n&=&E_r\,\hat{n^r}=\left[E_o \,\left(\vec{I}_{(+)a}
\hat{n}^a\right)+E_\ell\,\hat{K}_a\hat{n}^a\right]\,
\exp\,\left[i {{\omega}\over c}\,\Phi\right],\nonumber\\
&&\nonumber\\
E_{(\pm)}&=&E_r\,\epsilon^r_{(\mp)}=\left[E_o \,
\left(\vec{I}_{(+)a}
{\epsilon}^a_{(\mp)}\right)+E_\ell\,\hat{K}_a\,\epsilon^a_{(\pm)}\right]\,
\exp\,\left[{i\over c}\, \left(\mp \Omega\, \tau + \omega\,
\Phi(\tau,\vec{\sigma})\right)\right].
 \label{6.27}
 \eea

\bigskip

In the case $\hat{n}^a = \hat{K}^a$  we find

 \begin{eqnarray*}
B_n&=&B_{(-)}=0\nonumber\\
&&\nonumber\\
B_{(+)}&=&B_o \,e^{\,\left[{i\over c}\, \left( (\omega-\Omega) \tau
+ \vec{K}\cdot\vec{\sigma}\right)\right]},
 \end{eqnarray*}

 \bea
E_n&=&E_\ell\, e^{\,\left[i {{\omega}\over c}\, \left(
\omega \tau + \vec{K}\cdot\vec{\sigma}\right)\right]},\qquad E_{(+)}=0\nonumber\\
&&\nonumber\\
E_{(+)}&=&E_o\, e^{\,\left[{i\over c}\, \left( (\omega-\Omega) \tau
+ \vec{K}\cdot\vec{\sigma}\right)\right]}.
 \label{6.28}
 \eea

Therefore the components $B_{(+)},E_{(+)}$ have the frequency
modified to $\omega \mapsto \omega - \Omega$: this is the phase wrap
up effect. These are same results as in Ref.\cite{31} at the lowest
order in $\Omega/c$. The only new fact is the presence of the
component $E_n\neq 0$.

It would be interesting to make the calculation of the deviations of
order $O(\Omega^2/c^2)$ from rigid rotation, to see whether the
result $\omega \mapsto \gamma\, (\omega \pm \Omega)$ ($\gamma$ is a
Lorentz factor), found in Ref.\cite{31} by using the locality
hypothesis and supporting the interpretation with the
helicity-rotation coupling, is confirmed.

\subsection{The Sagnac Effect}

Following a suggestion of Ref.\cite{37} let us consider the solution
(\ref{6.8}) in the eikonal approximation, which describes the
propagation of the radiation along a ray of light whose trajectory
is given in Eq.(\ref{6.19}). This solution allows to get a
derivation of the Sagnac effect (described in Section II) along the
lines of Ref.\cite{31a}.
\medskip

Let us consider two receivers $A$ and $B$ at rest in the rotating
frame and characterized by the 3-coordinates $\eta^r_A$ and
$\eta^r_B$ respectively. Let us assume that $A$ and $B$ lie in the
same 2-plane containing the origin $\sigma^r = 0$ and orthogonal  to
$\vec \Omega$. Therefore we have $\vec{\Omega} \cdot \vec{\eta}_A =
\vec{\Omega} \cdot \vec{\eta}_B=0$. Let us assume that $A$ and $B$
are both on the trajectory of a ray of light, so that
Eq.(\ref{6.19}) implies the existence of a time $\tau_{AB}$ such
that we have

 \beq
\vec{\eta}_B-\vec{\eta}_A=\hat{k}\,\tau_{AB}+
\left(\frac{\vec{\Omega}}{c}\times\hat{k}\right)\,\tau^2_{AB}+{\cal
O}(\Omega^2/c^2).
 \label{6.29}
  \eeq
\medskip

The phase difference between $A$ and $B$ at the same instant $\tau$
is

 \bea
 \Delta\varphi_{AB}&=&\frac{\omega}{c}\, \left[\Phi(\tau,\vec{\eta}_B)
- \Phi(\tau,\vec{\eta}_A)\right] =\nonumber\\
&&\nonumber\\
&=&- \frac{\omega}{c}\, \left[\hat{k} \cdot (\vec{\eta}_B -
\vec{\eta}_A) + (\hat{k} \cdot \vec{\eta}_B)\,
\left(\frac{\vec{\Omega}}{c} \cdot \vec{\eta}_B \times
\hat{k}\right) - (\hat{k} \cdot \vec{\eta}_A)\,
\left(\frac{\vec{\Omega}}{c} \cdot \vec{\eta}_A \times
\hat{k}\right)\right] +\nonumber\\
&&\nonumber\\
&+&O(\Omega^2/c^2).
 \label{6.30}
 \eea

Eq.(\ref{6.29}) implies

 \beq
\vec{\eta}_B = \vec{\eta}_A + \hat{k}\, \tau_{AB} + O(\Omega/c)\,
\Rightarrow\, \frac{\vec{\Omega}}{c} \cdot \vec{\eta}_B \times
\hat{k} = \frac{\vec{\Omega}}{c} \cdot \vec{\eta}_A \times \hat{k} +
O(\Omega^2/c^2),
 \label{6.31}
 \eeq

\noindent so that we get

 \beq
\Delta\varphi_{AB} = - \frac{\omega}{c}\, \left[\hat{k} \cdot
(\vec{\eta}_B - \vec{\eta}_A) + \hat{k} \cdot (\vec{\eta}_B -
\vec{\eta}_A)\, \left(\frac{\vec{\Omega}}{c} \cdot \vec{\eta}_A
\times \hat{k}\right) + O(\Omega^2/c^2) \right].
 \label{6.32}
 \eeq
\medskip

Since  Eq.(\ref{6.29}) also implies $\vec{\eta}_B - \vec{\eta}_A =
\mid\vec{\eta}_B - \vec{\eta}_A\mid\, \hat{k} + O(\Omega^2/c^2)$, we
arrive at the result

 \beq
\Delta\varphi_{AB} = - \frac{\omega}{c}\, \left[\mid\vec{\eta}_B -
\vec{\eta}_A\mid + \frac{\vec{\Omega}}{c} \cdot \vec{\eta}_A \times
(\vec{\eta}_B - \vec{\eta}_A) \right] + O(\Omega^2/c^2).
 \label{6.33}
 \eeq

\medskip

If $A_{BAO}$ is the area of the triangle BAO in the 2-plane
orthogonal to $\vec \Omega$, we have $\frac{\vec{\Omega}}{c} \cdot
\vec{\eta}_A \times (\vec{\eta}_B - \vec{\eta}_A) = \pm 2\,
\frac{\Omega}{c}\, A_{BAO}$ (the choice of $\pm$ depends on the
direction of motion of the ray). As a consequence, the phase
difference is the sum of the following two terms

 \beq
\Delta\varphi_{AB} = - \frac{\omega}{c} \mid\vec{\eta}_B -
\vec{\eta}_A\mid + \delta\varphi_{AB} + O(\Omega^2/c^2).
 \label{6.34}
 \eeq

While the first term, $- \frac{\omega}{c} \mid \vec{\eta}_B -
\vec{\eta}_A\mid$, is present also in the inertial frames, the
second term

 \beq
\delta\varphi_{AB} = \mp \frac{2\, \omega\, \Omega}{c^2}\, A_{BAO},
 \label{6.35}
 \eeq

\noindent is the extra phase variation due to the rotation of the
frame. This is the Sagnac effect.

\subsection{The Inertial Faraday Rotation}

Let us give the derivation of the rotation of the polarization of an
electro-magnetic wave in a rotating frame, named inertial Faraday
rotation, which is important in astrophysics \cite{38}, were it is
induced by the gravitational field (due to the equivalence principle
only non-inertial frames are allowed in general relativity). Our
approach is analogous to the one of Ref.\cite{29} in the case of
Post-Newtonian gravity.
\bigskip

Let us consider the amplitude $\vec{a}$ of the solution (\ref{6.8})
in the eikonal approximation: it carries the information about the
polarization of a ray of light. To study the first-order transport
equation for it, the second of Eqs.(\ref{6.9}), let us make the
series expansion

 \beq
\vec{a}(\tau,\vec{\sigma}) = \vec{a}_o(\tau,\vec{\sigma}) +
\frac{\Omega}{c}\, \vec{a}_1(\tau,\vec{\sigma}) +
O\Big(\frac{\Omega^2}{c^2}\Big),
 \label{6.36}
 \eeq

\noindent and let us make the ansatz (in an inertial frame it
corresponds to a plane wave)

 \beq
\vec{a}_o(\tau,\vec{\sigma}) = \vec{a}_o = \mbox{ const.}, \quad
\Rightarrow \quad \frac{\partial\vec{a_o}}{\partial\tau} = 0,\qquad
\partial_r\,\vec{a}_o = 0.
 \label{6.37}
 \eeq
\medskip

This ansatz implies the following form of the second and third
equation in Eqs.(\ref{6.9})

 \bea
  &&\frac{\Omega}{c}\, \left[
\left(\frac{\partial\vec{a}_1}{\partial\tau} + \hat{\Omega} \times
\vec{a}_o\right) - (\hat{k} \cdot \vec{\partial})\, \vec{a}_1
\right] + O\Big(\frac{\Omega^2}{c^2}\Big) = 0,\nonumber\\
&&\nonumber\\
&&\vec{a}_o \cdot \hat{k} + \frac{\Omega}{c}\, \left[\vec{a}_o \cdot
\Big(\,\hat{k}\, (\hat{\Omega} \cdot \vec{\sigma} \times \hat{k}) -
(\hat{k} \cdot \vec{\sigma})\, (\hat{\Omega} \times \hat{k}) \,\Big)
 + \vec{a}_1 \cdot \hat{k}\right] + O\Big(\frac{\Omega^2}{c^2}\Big) = 0.
 \label{6.38}
 \eea

To study these equations, let us assume that each rotating receiver
is endowed with a tetrad of the type given in Eq.(\ref{6.20}): the
spatial axes ${\tilde {\cal W}}^r_{(i)} = (R^r_{1}(k), R^r_{2}(k),
\hat{k}^r)$ with $\vec{R}_{\lambda}(k) \cdot \vec{R}_{\lambda'}(k) =
\delta_{\lambda\lambda'}$, $\vec{R}_\lambda(k) \cdot \hat{k} = 0$.
\medskip

The second of Eqs.(\ref{6.38}) for the unknown $\vec{a}_o$,
$\vec{a}_1$ is the transversality condition and it implies

\bea
 order\,\,\, 0\,\,\, in\,\,\, \Omega &\rightarrow& \vec{a}_o \cdot \hat{k}
 = 0\, \Rightarrow\, \vec{a}_o = a_o^\lambda\, \vec{R}_\lambda(k),
 \nonumber \\
 order\,\,\, 1\,\,\, in\,\,\, \Omega &&\nonumber \\
 &&{}\nonumber \\
 \vec{a}_1 \cdot \hat{k}&=&- \vec{a}_o \cdot \Big(\,\hat{k}\, (\hat{\Omega}
 \cdot \vec{\sigma} \times \hat{k}) + (\hat{k} \cdot \vec{\sigma})\,
 (\hat{\Omega} \times \hat{k})\,\Big)=\nonumber\\
&&\nonumber\\
&=&- a_o^\lambda\, \vec{R}_\lambda(k) \cdot (\hat{\Omega} \times
\hat{k})\, (\hat{k} \cdot \vec{\sigma}).
 \label{6.39}
 \eea

\medskip

Due to the ansatz (\ref{6.37}) the first of Eqs.(\ref{6.38}) is of
order 1 in $\Omega$ and gives the following condition on $\vec{a}_1$

 \beq
\frac{\partial\vec{a}_1}{\partial\tau} - \hat{\Omega} \times
\vec{a}_o + (\hat{k} \cdot \vec{\partial})\, \vec{a}_1 = 0.
 \label{6.40}
 \eeq

If we project this equation on the directions $\hat{k}$,
$\vec{R}_\lambda(k)$, we get

 \bea
&&\frac{\partial}{\partial\tau}\, (\vec{a_1} \cdot \hat{k}) -
\hat{\Omega} \times \vec{a}_o \cdot \hat{k} + (\hat{k} \cdot
\vec{\partial})\, (\vec{a}_1 \cdot \hat{k}) = 0,\nonumber\\
&&\nonumber\\
&&\frac{\partial a_1^\lambda}{\partial\tau} - \hat{\Omega} \times
\vec{R}_{\lambda'}(k) \cdot \vec{R}_\lambda(k)\, a_o^{\lambda'} +
(\hat{k} \cdot \vec{\partial})\, a_1^\lambda = 0.
 \label{6.41}
 \eea

While the first of Eqs.(\ref{6.41}) is automatically satisfied, the
second one is an equation for the components $a_1^\lambda$. The
simplest solutions are obtained with the following ansatz

 \beq
\frac{\partial a_1^\lambda}{\partial\tau} = 0,\quad \Rightarrow
\quad a_1^\lambda(\tau) = \left[\hat{\Omega} \times
\vec{R}_{\lambda'}(k) \cdot\vec{R}_\lambda(k)\right]\, \hat{k} \cdot
\vec{\sigma}\, a_o^{\lambda'}.
 \label{6.42}
 \eeq

\medskip

The final solution for the transverse electro-magnetic potential is

 \bea
\vec{A}_\perp&=&\frac{a_o{}^1}{\omega}\, \left[\vec{R}_{1} +
\theta(\vec{\sigma})\, \vec{R}_{2}(k) - \frac{\hat{k} \cdot
\vec{\sigma}}{c}\, (\vec{\Omega} \cdot \vec{R}_2(k))\, \hat{k}
\right]\, e^{\,\left(\,i\, \frac{\omega}{c}\, \Phi\right)} +\nonumber\\
&&\nonumber\\
&+&\frac{a_o{}^2}{\omega}\, \left[\vec{R}_{2}(k) -
\theta(\vec{\sigma})\, \vec{R}_{1}(k) + \frac{\hat{k} \cdot
\vec{\sigma}}{c}\, (\vec{\Omega} \cdot \vec{R}_1(k))\, \hat{k}
\right]\, e^{\,\left(\,i\, \frac{\omega}{c}\, \Phi\right)} +
O(1/\omega^2),\nonumber \\
 &&{}\nonumber \\
 &&with\nonumber \\
 &&{}\nonumber \\
 \theta(\vec{\sigma}) &=& \frac{1}{c}\, (\hat{k} \cdot
\vec{\sigma})\, (\vec{\Omega} \cdot \hat{k}).
 \label{6.43}
 \eea
\medskip

The resulting non-inertial magnetic and electric fields are
($\vec{B} = \{B_r\}$, $\vec{E} = \{E_r\}$)

 \begin{eqnarray*}
  \vec{B}&=&- \frac{i\, a_o{}^1}{c}\, \left[\,\vec{R}_{2}(k)
  - \theta(\vec{\sigma})\, \vec{R}_{1}(k)\,\right]\,
e^{\,\left(\,i\, \frac{\omega}{c}\, \Phi\right)} -\nonumber\\
&&\nonumber\\
&-&\frac{i\, a_o{}^1}{c}\, \left[\left(\frac{\vec{\Omega}}{c} \times
\vec{\sigma} \cdot \hat{k}\right)\, \vec{R}_{2}(k) - \frac{\hat{k}
\cdot \vec{\sigma}}{c}\, (\vec{\Omega} \cdot \vec{R}_1(k))\,
\hat{k}\, \right]\, e^{\,\left(\,i\, \frac{\omega}{c}\, \Phi\right)}
+\nonumber\\
&&\nonumber\\
&+&\frac{i\, a_o{}^2}{c}\, \left[\,\vec{R}_{1}(k) +
\theta(\vec{\sigma})\, \vec{R}_{2}(k)\,\right]\,
e^{\,\left(\,i\, \frac{\omega}{c}\, \Phi\right)} +\nonumber\\
&&\nonumber\\
&+&\frac{i\, a_o{}^2}{c}\, \left[\left(\frac{\vec{\Omega}}{c} \times
\vec{\sigma} \cdot \hat{k}\right)\, \vec{R}_{1}(k) + \frac{\hat{k}
\cdot \vec{\sigma}}{c}\, (\vec{\Omega} \cdot \vec{R}_2(k))\,
\hat{k}\,
\right]\, e^{\,\left(\,i\, \frac{\omega}{c}\, \Phi\right)} +\nonumber\\
&&\nonumber\\
&+&O(1/\omega) + O(\Omega^2/c^2) =\nonumber\\
&&\nonumber\\
&\byd&\,b(\vec{\sigma})\,\, e^{\,\left(\,i\, \frac{\omega}{c}\,
\Phi\right)} + O(1/\omega) + O(\Omega^2/c^2),
\end{eqnarray*}

\bea
 \vec{E}&=&- \frac{i\, a_o{}^1}{c}\, \left[\vec{R}_{1}(k) + \theta(\vec{\sigma})\,
 \vec{R}_{2}(k) - \frac{\hat{k} \cdot \vec{\sigma}}{c}\, (\vec{\Omega}
 \cdot \vec{R}_2(k))\, \hat{k}\right]\, e^{\,\left(\,i\, \frac{\omega}{c}\,
 \Phi\right)} -\nonumber\\
&&\nonumber\\
&-&\frac{i\, a_o{}^2}{c}\, \left[\vec{R}_{2}(k) -
\theta(\vec{\sigma})\, \vec{R}_{1}(k) + \frac{\hat{k} \cdot
\vec{\sigma}}{c}\, (\vec{\Omega} \cdot \vec{R}_1(k))\, \hat{k}
\right]\, e^{\,\left(\,i\, \frac{\omega}{c}\, \Phi\right)} +\nonumber\\
&&\nonumber\\
&+&O(1/\omega) + O(\Omega^2/c^2) =\nonumber\\
&&\nonumber\\
&\byd&\,f(\vec{\sigma})\,\, e^{\,\left(\,i\, \frac{\omega}{c}\,
\Phi\right)} + O(1/\omega) + O(\Omega^2/c^2).
 \label{6.44}
 \eea

\medskip

As in the case of the Sagnac effect let us consider two receivers
$A$ and $B$ at the endpoints of the same light ray described by Eqs,
(\ref{6.19}) and (\ref{6.29}). The magnetic field observed by $A$,
$\vec{B}(\tau,\vec{\eta}_A)$, differs from the one observed by $B$,
$\vec{B}(\tau,\vec{\eta}_B)$. Since the phase changes have been
already analyzed for the Sagnac effect, let us concetrate on the
amplitudes $\vec{b}(\vec{\eta}_A)$ and $\vec{b}(\vec{\eta}_B)$.
Since Eq.(\ref{6.29}) gives $\vec{\eta}_B - \vec{\eta}_A = \hat{k}\,
\tau_{AB} + O(\Omega/c)$, we find

 \bea
  \vec{b}(\vec{\eta}_B) - \vec{b}(\vec{\eta}_A)&=&
\frac{i\, a_o{}^1}{c}\, \delta\theta_{BA}\, \vec{R}_{1}(k) +
\frac{i\, a_o{}^2}{c}\, \delta\theta_{BA}\, \vec{R}_{2}(k) +\nonumber\\
&&\nonumber\\
&+&\frac{i\, a_o{}^1}{c}\, \left[\frac{\mid\vec{\eta}_B -
\vec{\eta}_A\mid}{c}\, (\vec{\Omega} \cdot \vec{R}_1(k))\,
\hat{k}\,\right] +\nonumber\\
&&\nonumber\\
&+&\frac{i\, a_o{}^2}{c}\, \left[\frac{\mid\vec{\eta}_B -
\vec{\eta}_A\mid}{c}\, (\vec{\Omega} \cdot \vec{R}_2(k))\,
\hat{k}\,\right] + O(\Omega^2/c^2),\nonumber \\
 &&{}\nonumber \\
 &&with\nonumber \\
 &&{}\nonumber \\
 \delta\theta_{BA}&=&\theta(\vec{\eta}_B) - \theta(\vec{\eta}_A) =
\frac{1}{c}\, \mid\vec{\eta}_B - \vec{\eta}_A\mid\, (\vec{\Omega}
\cdot \hat{k}) + O(\Omega^2/c^2).
 \label{6.45}
 \eea

\noindent $\delta\theta_{BA}$ is the angle of the {\it inertial
Faraday rotation} (in this case it is small, $\delta\theta_{AB} \sim
\Omega/c$). It agrees with Eq.(4) of Ref.\cite{38}, where it has the
form $\delta\theta_{AB} = - \frac{1}{2}\, \int_A^B\,
\sqrt{g_{\tau\tau}}\, (\nabla \times \vec{n}) \cdot d\vec{\sigma}$
as a line integral along the spatial trajectory of the light ray.
This formula agrees with our result, because, due to the
approximations we have done, we have $g_{\tau\tau} = 1$, $(\nabla
\times \vec{n}) = - \frac{2\, \vec{\Omega}}{c}$ and our ray
trajectory is $\vec{\sigma}(\tau) = \hat{k}\, \tau + \vec{\sigma}_o
+ O(\Omega^2/c^2)$.

\medskip

To make the {\it rotation} explicit, let us write the components
along the two polarization directions: $b_{(\lambda)}(\vec{\eta}_A)
= \vec{b}(\vec{\eta}_A) \cdot \vec{R}_\lambda(k)$ and
$b_{(\lambda)}(\vec{\eta}_B) = \vec{b}(\vec{\eta}_B) \cdot
\vec{R}_\lambda(k)$. In this way we get

 \bea
b_{(1)}(\vec{\eta}_B)&=&b_{(1)}(\vec{\eta}_A) + \delta\theta_{AB}\,
\frac{i\, a_o{}^1}{c} + O(\Omega^2/c^2) = b_{(1)}(\vec{\eta}_A) -
\delta\theta_{AB}\, b_{(2)}(\vec{\eta}_A) + O(\Omega^2/c^2),\nonumber\\
&&\nonumber\\
b_{(2)}(\vec{\eta}_B)&=&b_{(2)}(\vec{\eta}_A) + \delta\theta_{AB}\,
\frac{i\, a_o{}^2}{c} + O(\Omega^2/c^2) = b_{(2)}(\vec{\eta}_A) +
\delta\theta_{AB}\, b_{(1)}(\vec{\eta}_A) + O(\Omega^2/c^2).
\nonumber \\
&&{}
 \label{6.46}
 \eea

\noindent This is just a small angle rotation with
$b_{(\lambda)}(\vec{\eta}_B) = R_{\lambda}{}^{\lambda'}(k)\,
(\delta\theta_{AB})\, b_{(\lambda')}(\vec{\eta}_A)$.

\medskip

The electric field may be treated in the same way.

\vfill\eject

\section{Conclusions}

The theory of non-inertial frames developed in these two papers is
free by construction from the coordinate singularities of all the
approaches to accelerated frames based on the 1+3 point of view, in
which the instantaneous 3-spaces are identified with the local rest
frames of the observer. The pathologies of this approach are either
the horizon problem of the rotating disk (rotational velocities
higher than $c$), which is still present in all the calculations of
pulsar magnetosphere in the form of the light cylinder, or the
intersection of the local rest 3-spaces. The main difference between
the 3+1 and 1+3 points of view is that the M$\o$ller conditions
forbid {\it rigid rotations} in relativistic theories.
\bigskip

In this paper we have given the simplest example of 3+1 splitting
with {\it differential rotations} and we have revisited the rotating
disk and the Sagnac effect following the 3+1 point of view. This
splitting is also used to give a special relativistic generalization
of the non-relativistic non-inertial International Terrestrial
Reference System (ITRS) used to describe fixed coordinates on the
surface of the rotating Earth in the conventions IERS2003 \cite{39}.

\bigskip

Then we re-examined some properties of the electro-magnetic wave
solutions of non-inertial Maxwell equations, which till now were
described only by means of the 1+3 point of view, in the 3+1
framework, where there is a well-posed Cauchy problem due to the
absence of coordinate singularities. By considering admissible
nearly rigid rotating frames we recover the results of the 1+3
approach and open the possibility to make these calculations in
presence of deviations from rigid rotations.

\bigskip

A still open problem are the constitutive equations for
electrodynamics in material media in non-inertial systems. For
linear isotropic media see the Wilson-Wilson experiment in
Refs.\cite{40} and Refs.\cite{37,41}, while for an attempt towards a
general theory in arbitrary media (including the premetric extension
of electro-magnetism) see Refs.\cite{42}

\bigskip

In conclusion we have now a good understanding of particles and
electro-magnetism in non-inertial frames in Minkowski space-time,
where the 4-metric induced by the admissible 3+1 splitting describes
all the inertial effects. Going to canonical gravity, in
asymptotically Minkowskian space-times without super-translations
and in the York canonical basis of Refs.\cite{12,a}, it is possible
to see which components remain inertial effects and which become
dynamical tidal effects (the physical degrees of freedom of the
gravitational field). Moreover the inertial 3-volume element and
some inertial components of the extrinsic curvature of the
instantaneous 3-spaces become complicated functions of both general
relativistic inertial and tidal effects, because they are determined
by the solution of the super-Hamiltonian constraint (the
Lichnerowicz equation) and of the super-momentum constraints.
Finally, in accord with the equivalence principle, the instantaneous
3-spaces are only partially determined by the freedom in choosing
the convention for clock synchronization: after such a convention
the final instantaneous 3-spaces associated to each solution of
Einstein's equations are dynamically determined, because in general
relativity the metric structure of space-time is dynamical and not
absolute like it happens in special relativity.

 \vfill\eject

\end{document}